\documentclass[11pt,a4paper]{article}

\usepackage{graphicx,array}
\usepackage{color}
\usepackage{latexsym}
\usepackage{amsthm}
\usepackage{amsmath}
\usepackage{amssymb}
\usepackage{siunitx}
\usepackage{empheq}
\usepackage{braket}
\usepackage[version=4]{mhchem}
\usepackage{bm}
\usepackage{bbm}
\usepackage{dsfont}
\usepackage{epsfig}
\usepackage{slashed}
\usepackage{bbold}
\usepackage{psfrag}
\usepackage[svgnames]{xcolor}
\PassOptionsToPackage{caption=false}{subfig}
\usepackage{subcaption}
\usepackage{xfrac}
\usepackage{multirow}
\usepackage{booktabs}
\usepackage[normalem]{ulem}
\usepackage{jheppub}
\usepackage{adjustbox}
\usepackage{mathtools}
\usepackage{indentfirst}

\def\bl{\begin{equation}\begin{aligned}}
\def\el{\end{aligned}\end{equation}}
\def\beal{\begin{align}}
\def\eal{\end{align}}
\def\be{\begin{equation}}
\def\ee{\end{equation}}
\def\bpm{\begin{pmatrix}}
\def\epm{\end{pmatrix}}
\def\bsm{\begin{bmatrix}}
\def\esm{\end{bmatrix}}
\def\bvm{\begin{vmatrix}}
\def\evm{\end{vmatrix}}
\def\bVM{\begin{Vmatrix}}
\def\eVM{\end{Vmatrix}}
\def\bea{\begin{eqnarray}}
\def\eea{\end{eqnarray}}

\newcommand{\lan}{\langle}

\def\1{{\bf 1}}
\def\2{{\bf 2}}
\def\3{{\bf 3}}
\def\4{{\bf 4}}
 
 \def\SU{{\mathrm{SU}}}

\newcommand{\bg}[1]{\textcolor{blue}{ #1}}

\title{Tetraquarks in the Born-Oppenheimer approximation}
\date{\today}
\author[a]{D. Germani,}\emailAdd{davide.germani@uniroma1.it}
\author[b]{B. Grinstein,}
\author[a]{A.D. Polosa}
\affiliation[a]{
Universit\`a degli Studi di Roma La Sapienza and INFN Section of Roma 1, Piazzale Aldo Moro 5, 00185 Roma, Italy
}
\affiliation[b]{
University of California, San Diego, 9500 Gilman Drive, La Jolla, CA 92093, USA
}

\abstract{
The conventional loosely bound molecule interpretation of the $X(3872)$ is not compatible with the recent LHCb experimental measurement of the ratio of branching fractions  $\mathcal{R}=\text{Br}(X\to\psi^\prime\gamma)/\text{Br}(X\to\psi\gamma)$. We systematically  determine the entire tetraquark spectrum for $J=0,1,2$ and refine the calculation of  $\mathcal{R}$ in an improved Born-Oppenheimer description of the $X(3872)$ compact tetraquark.
This refinement yields a significantly better agreement with experimental data on ${\cal R}$ and on the spectroscopy of the states themselves.   
Extending the diquark-antidiquark paradigm to encompass tetraquarks that are linear superpositions of open charm singlets and color octets, we discover that these exotic resonances manifest as compact shallow bound states of quarks in color force potentials.}

\begin{document}
\maketitle

\newpage

\section{Introduction}
The degree of compositeness of the $X(3872)$ remains a subject of ongoing debate, with various theoretical possibilities being considered. These include conventional $c\bar c$ or exotic tetraquark compact states, as well as the mesonic open charm molecules. In the molecular model, the $X$ is described as a shallow bound state of a $D$ and a $\bar{D}^*$ meson \cite{molecule1,molecule2,molecule3,molecule4,molecule5}. In the compact model, quarks are bound by QCD interactions \cite{compact1,compact2,hbondqcd1,hbondqcd2,hbondqcd3,x3872,nora,lattice_Braaten}, while in \cite{DeFazio:2008xq,cc2,cc3andradiative,Colangelo:2025uhs}, the conventional $c\bar{c}$ charmonium hypothesis is investigated. In~\cite{rr4q}, a qualitative discussion on compact tetraquarks $Q\bar{Q}q\bar{q}$ is provided in the large-$N$ limit, along with a more in-depth analysis of $QQ\bar{q}\bar{q}$ tetraquarks.

The study of radiative decays of the $X(3872)$ into $\psi^\prime$ or $J/\psi$ provides valuable insights into its structure. A precise measurement of the ratio of branching fractions $\mathcal{R}=\text{Br}(X\to\psi^\prime\gamma)/\text{Br}(X\to J/\psi\gamma)$ was recently achieved by the LHCb collaboration \cite{LHCb}
\begin{equation}
    \mathcal{R}_{\text{exp}}=1.67\pm0.21\pm0.12\pm0.04.
\end{equation}
This value conflicts with the conventional interpretation of the $X$ as a $D\bar D^*$ molecule with a relatively low binding energy. In non-relativistic scattering theory, shallow bound states are characterized by a universal wavefunction that is independent of the specific details of the (unknown) binding potential \cite{lqm3-2}. In \cite{molecradiative1,x3872}, $\mathcal{R}$ is calculated under the molecular hypothesis, yielding values significantly less than unity. 
This is a clear indication that LHCb’s result cannot be reproduced solely based on the universal wavefunction of a loosely bound $D\bar D^*$ state. One can accommodate the observed value of the ratio $\mathcal{R}$ in the molecular model as described by the hadronic EFT techniques which incorporate additional tunable parameters \cite{Guo:2014taa,Molnar:2016dbo}. As outlined in \cite{ccradiative1,ccradiative2}, the $P$-wave charmonium hypothesis gives $4\leq \mathcal{R}\leq 6$ whereas in \cite{cc3andradiative} it is found $1<\mathcal{R}\lesssim 10^5$. In a recent analysis \cite{Colangelo:2025uhs}, $\mathcal{R} = 1.7 \pm 0.3$ is found. The compact tetraquark hypothesis, as proposed in \cite{x3872, nora}, exhibits excellent compatibility with the experimentally measured $\mathcal{R}_{\text{exp}}$. A summary of theoretical predictions for $\mathcal{R}$ can be found in \cite{LHCb}.

In this paper, we introduce an improved version of the Born-Oppenheimer (BO) description of the compact tetraquark $c\bar{c}q\bar{q}$ employed in \cite{x3872}. The compact tetraquark in \cite{x3872} comprises only diquark-antidiquark, $cq$ and $\bar c\bar q$, “orbitals,” and lacks spin interactions. In contrast, this study incorporates both $cq$ and $\bar c q$ orbitals, as is customary in the context of heteronuclear molecules in quantum mechanics, utilizing the linear combination of atomic orbitals (LCAO) approximation \cite{bj,Pauling}. By incorporating the spin potential, we compute the spectrum for tetraquarks from $J=0$ to $J=2$ and compare our results with the experimental observations and measurements collected in the PDG \cite{pdg}.

We perform an updated calculation, with respect to~\cite{x3872}, of the compact ratio $\mathcal{R}_{\text{comp}}= 1.4\pm0.3$ as illustrated in Sec.~\ref{sec:rad}. 
In addition we gain a qualitatively new understanding of tetraquarks in the BO approximation: at large separations between the heavy quarks, the lowest energy compact tetraquarks are found to be superpositions of meson-antimeson and $(c\bar q)_{\bm 8}(c\bar q)_{\bm 8}$ color states.  
 
\section{Born-Oppenheimer tetraquarks}
\label{sec:BOT}
\begin{figure}[t]
    \centering
    \includegraphics[width=0.45\textwidth]{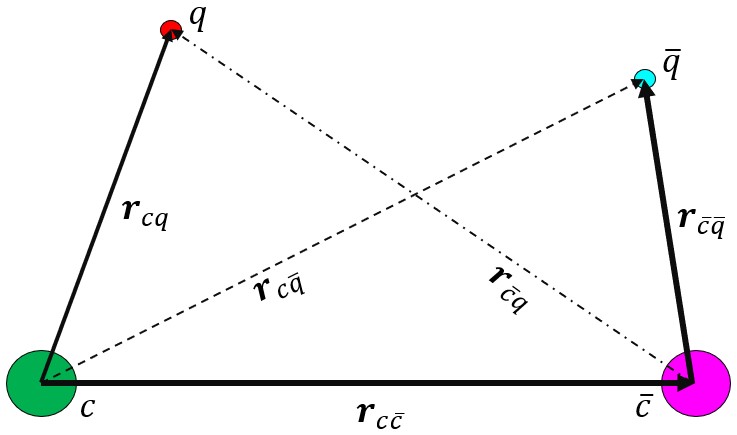}
    \caption{Coordinate system adopted in the Born-Oppenheimer approximation. The coordinates satisfy the relations: $\bm{r}_{c\bar{q}}=\bm{r}_{\bar{c}\bar{q}}+\bm{r}_{c\bar{c}}$ and $\bm{r}_{\bar{c}q}=\bm{r}_{cq}-\bm{r}_{c\bar{c}}$.} 
    \label{fig:Bocoord}
\end{figure}

In the  BO approximation scheme  the wave function of the tetraquark, $\Psi_T$, can be factorized by the product of a wave function for the fast degrees of freedom (the light quarks) $\Psi_{q\bar{q}}$ times the slow degrees of freedom (the heavy quarks) $\Psi_{c\bar{c}}$. Using the notation in Fig.~\ref{fig:Bocoord}, this factorization is expressed by
\begin{equation}
    \Psi_T(\bm{r}_{cq},\bm{r}_{\bar{c}\bar{q}},\bm{r}_{c\bar{c}})=\Psi_{c\bar{c}}(\bm{r}_{c\bar{c}})\Psi_{q\bar{q}}(\bm{r}_{cq},\bm{r}_{\bar{c}\bar{q}},\bm{r}_{c\bar{c}})\,.
\end{equation}
The function $\Psi_{q\bar{q}}$ is the solution to the Schr\"odinger equation
\begin{equation}
    H_{q\bar{q}}\Psi_{q\bar{q}}(\bm{r}_{cq},\bm{r}_{\bar{c}\bar{q}},\bm{r}_{c\bar{c}})=\Delta E(r_{c\bar{c}})\,\Psi_{q\bar{q}}(\bm{r}_{cq},\bm{r}_{\bar{c}\bar{q}},\bm{r}_{c\bar{c}})\,,
    \label{lqp}
\end{equation}
where the hamiltonian $H_{q\bar{q}}$ includes all the interactions involving the light quarks, and the coordinate $r_{c\bar{c}}$ enters parametrically. Once $\Delta E(r_{c\bar{c}})$ is obtained, the function $\Psi_{c\bar{c}}$ is the solution of
\begin{equation}
    \left(H_{c\bar{c}}+\Delta E(r_{c\bar{c}})\right)\Psi_{c\bar{c}}(\bm{r}_{c\bar{c}})=E\,\Psi_{c\bar{c}}(\bm{r}_{c\bar{c}})\,,
\end{equation}
where the Hamiltonian $H_{c\bar{c}}$ describes the kinetic energy of, and the interactions between, the heavy quarks, and the effect of the interactions among  the moving  light quarks and between them and the static charm quarks is captured by the potential $\Delta E(r_{c\bar{c}})$.

The light quark problem~\eqref{lqp} cannot be solved analytically. We will seek approximate expressions for $\Psi_{q\bar{q}}$ and $\Delta E(r_{c\bar{c}})$ using  variational methods. Before proceeding, we list the parameters to be used from now on. The constituent quark masses $m_q$ and $m_c$, the strong coupling constant $\alpha_s$, and the string tension between heavy-heavy and heavy-light quarks $k$ are given by \cite{charmspetr,multiquark}
\begin{equation}
    m_q=308\,\text{MeV}\,,\quad\quad m_c=1317\,\text{MeV} \,,\quad\quad \alpha_s(2M_c)=0.331\,, \quad\quad k=0.176\,\text{GeV}^2.
    \label{eq:parametri}
\end{equation}
The mass of the charm quark and the string tension $k$ are chosen so that the $c\bar{c}$ Cornell potential 
\begin{equation}
    V(r_{c\bar{c}})=-\frac{4}{3}\alpha_s \frac{1}{r_{c\bar{c}}}+k\,r_{c\bar{c}}
\end{equation}
provides a  mass difference $\psi(2S)-\psi(1S)$  equal to the  measured one.

There are two ways of combining two quarks and two anti-quarks into a colorless state. The compact hypothesis used in this paper, henceforth ``compact tetraqurk'', assumes that the invariant state $T$ is formed by combining first the $c\bar c$ pair in a color octet, thus
\be
T=|(c\bar c)_{\bm 8}(q\bar q)_{\bm 8}\rangle^{\phantom{\dagger}}_{\bm 1}
\ee
The subscript ``$\bm1  $'' indicates that the state is a color singlet; below we omit it when it is clear that the context involves colorless states. Alternatively, one can  present this state as a fixed linear combinations of two independent singlets obtained by first combining the $cq$ pair or the $c\bar q$ pair, into irreducible representations, which we will call $D$ (diquarks) and $M$ (meson and adjoint meson)
\be
T_D=\sqrt{\frac{2}{3}}|(cq)_{\boldsymbol{\bar{3}}}(\bar{c}\bar{q})_{\boldsymbol{3}}\rangle-\sqrt{\frac{1}{3}}|(cq)_{\boldsymbol{6}}(\bar{c}\bar{q})_{\boldsymbol{\bar{6}}}\rangle
\label{32}
\ee
and 
\be
T_M=\sqrt{\frac{8}{9}}|(c\bar{q})_{\boldsymbol{1}}(\bar{c}q)_{\boldsymbol{1}}\rangle-\sqrt{\frac{1}{9}}|(c\bar{q})_{\boldsymbol{8}}(\bar{c}q)_{\boldsymbol{8}}\rangle
\label{33}
\ee
The fixed coefficients are readily computed using color-Fierz rearrangements of $T$. Physically this means, for example, from~\eqref{32},  that the $(cq)(\bar{c}\bar{q})$ pairs appear with probability $2/3$ in the color configuration $\bar{\bm 3},\bm 3$ and with probability $1/3$ in $\bm 6, \bar{\bm 6}$.

At short distances, the dominant force between the quarks is Coulomb-like, and originates in one gluon-exchange. Its strength depends on the color configuration of participating quarks. 
Consider, for example, the $cq$ pair exchanging one gluon in the $t$-channel. The color factor is $T^a_{iI}T^a_{jJ}$, sum on $a$ implicit, where $i,j$ are the colors of the incoming $cq$ pair and $IJ$ are the colors of the outgoing $cq$ pair. The $cq$ pair can be  either in $\bar{\bm 3}$-color, or in $\bm 6$, and we aim at computing the interaction strength for each configuration. The method is well known. First, solve for $T^a_{iI}T^a_{jJ}$ in
\begin{equation}
  \label{eq:TaTa}
\begin{aligned}
    (T^a_{im}\delta_{jn}+\delta_{im}T^a_{jn})
  (T^a_{mI}\delta_{nJ}+\delta_{mI}T^a_{nJ})
  &=(T^aT^a)_{iI}\delta_{jJ}+ \delta_{iI}(T^aT^a)_{jJ}+ 2 T^a_{iI}T^a_{jJ}
  \\
  &=C(R_1)\delta_{iI}\delta_{jJ} + C(R_2)\delta_{iI}\delta_{jJ} + 2 T^a_{iI}T^a_{jJ}
  \end{aligned}
\end{equation}   
In the second line $C(R)$ stands for  the quadratic Casimir invariant for $R$ representation, and we have kept general the representation $R_1$ and $R_2$ of the two quarks, for now. The left hand side of \eqref{eq:TaTa} has the Casimir of the tensor product representation, which can be expressed as the direct sum 
\be
\bigoplus_k C(S_k)\,  I_{D(S_k)}
\ee
where $I_{D(S_k)}$ is the identity matrix of dimension $D(S_k)$, and the $S_k$ are  the irreducible components in the direct sum
\be
R_1\otimes R_2 = S_1\oplus S_2\oplus S_3\dots 
\ee
With this, one obtains   that, when acting on a pair of quarks in the color representation $S$,  $T^a_{iI}T^a_{jJ}=\lambda(S) \delta_{iI}\delta_{jJ}$  where
\begin{equation}
\lambda(S)=\tfrac{1}{2}(C(S)-C(R_1)-C(R_2))
\label{eq:casimir}
\end{equation}
For example, from~\eqref{32} (the $D$ Fierzing),   we have
\be
\lambda_{cq}=\lambda_{\bar c\bar q}=\frac{2}{3}\left(\frac{1}{2}\left(\frac{4}{3}- \frac{8}{3}\right)\right)+\frac{1}{3}\left(\frac{1}{2}\left(\frac{10}{3}- \frac{8}{3}\right)\right)=-\frac{1}{3}
\ee
since $C(\bm 3)=C(\bar{\bm 3})=4/3$ and $C(\bm 6)=C(\bar{\bm 6})=10/3$. 
We get similarly
\be
\lambda_{c\bar{c}}=\lambda_{q\bar{q}}=
+\frac{1}{6}
\ee
and 
\be
\lambda_{c\bar{q}}=\lambda_{\bar{c}q}=-\frac{7}{6}
\label{eq:lambda_cbarq}
\ee

According to these results, the $cq$ potential $V_D$ and the $c\bar q$ potential  $V_M$ are given by
\be
V_{D}(\bm{\zeta})=-\frac{1}{3}\alpha_s\frac{1}{\zeta}+k\,\zeta
\label{eq:VA}
\ee
where $\bm \zeta=\bm{r}_{cq},\,\bm{r}_{\bar{c}\bar{q}}$ and 
\be
V_{M}(\bm{\zeta})=-\frac{7}{6}\alpha_s\frac{1}{\zeta}+k\,\zeta
\label{eq:VB}
\ee
where $\bm \zeta=\bm{r}_{c\bar{q}},\,\bm{r}_{\bar{c}q}\,$. 

To compute the wavefunctions associated to the potentials \eqref{eq:VA} and \eqref{eq:VB}, we introduce the variational test-functions, in the hydrogen-like form
\be
\psi_{\cal C}(\bm \zeta)=\sqrt{\frac{{\cal C}^3}{\pi}}e^{-{\cal C}\zeta}\qquad\text{where}\qquad  {\cal C}=D,M
\label{varfunc}
\ee
and we calculate $\cal{C}$ by minimizing the Hamiltonians containing $V_D$, $V_M$ and the respective kinetic terms. The mean value can be computed analytically for both potentials by minimizing
\begin{equation}
E_\mathcal{C}=\left\lan\psi_{\cal{C}}(\bm{\zeta})\left|\,-\frac{\nabla^2}{2m_{cq}}+V_{\cal C}(\bm{\zeta})\,\right|\psi_{\cal{C}}(\bm{\zeta})\right\rangle=\frac{\mathcal{C}^2}{2m_{cq}}+\lambda_{\cal C}\alpha_s\,{\cal C}+\frac{3k}{2\cal{C}}\qquad {\cal C}=D,M
    \label{eq:E_C}
\end{equation}
where $m_{cq}$ is the reduced mass of the  $cq$ system and $\lambda_{\cal C}$ is $-1/3$ or $-7/6$.\footnote{The integrals involved in each of the three terms on the right hand side of~\eqref{eq:E_C} correspond to $I_0^{\mathcal{C}}$, $I_1^{\mathcal{C}}(0)$ and $J_2^{\mathcal{C}}$ of Eqs.~\eqref{app:I0A}, ~\eqref{app:I1A} and~~\eqref{app:J2A}, respectively. }  With a slight abuse of notation, depending on where  ${\cal C}=D,M$ appears it can denote a label, as in the case of the subscripts in~\eqref{eq:E_C},  or   numerical values  if it appears as a parameter in a formula, as on the right hand side (rhs) of~\eqref{eq:E_C}. We find 
\begin{equation}
D=0.413\,\text{GeV} \qquad \langle E_{D}\rangle^{\phantom{\dagger}}_{\rm min}=0.935\,\text{GeV}
\label{eq:E_A}
\end{equation}
and
\be
 M=0.439\,\text{GeV} \qquad \langle E_{M}\rangle^{\phantom{\dagger}}_{\rm min}=0.818\,\text{GeV.}
 \label{eq:E_B}
\ee

We  factorize $\Psi_{q\bar q}$ in~\eqref{lqp} in terms of the functions $\psi_{\cal C}$ in~\eqref{varfunc} as in 
\begin{equation}
\Psi_{q\bar{q}}(\bm{r}_{cq},\bm{r}_{\bar{c}\bar{q}},\bm{r}_{c\bar{c}})=a_1(r_{c\bar{c}})\,\psi_D(\bm{r}_{cq})\psi_D(\bm{r}_{\bar{c}\bar{q}})\, +\, a_2(r_{c\bar{c}})\,\psi_M(\bm{r}_{c\bar{q}})\psi_M(\bm{r}_{\bar{c}q})\,;
 \label{41}
\end{equation}
see Fig.~\ref{fig:Bocoord}. 
The case $a_1(r_{c\bar{c}})=1,a_2(r_{c\bar{c}})=0$~\cite{x3872} corresponds to
\begin{equation}
\Psi_{q\bar q}(\bm{r}_{cq},\bm{r}_{\bar{c}\bar{q}},\bm{r}_{c\bar{c}})=\psi_D(\bm{r}_{cq})\psi_D(\bm{r}_{\bar{c}\bar{q}})\qquad \bm 3,\bar{\bm 3}+\bm 6,\bar{\bm 6}
\label{42}
\end{equation} 
 whereas $a_1(r_{c\bar{c}})=0,\,a_2(r_{c\bar{c}})=1$ corresponds to
\begin{equation}
\Psi_{q\bar q}(\bm{r}_{cq},\bm{r}_{\bar{c}\bar{q}},\bm{r}_{c\bar{c}})=\psi_M(\bm{r}_{c\bar{q}})\psi_M(\bm{r}_{\bar{c}q})\qquad \bm 1,\bm 1+{\bm 8,\bm 8}
\label{43}
\end{equation}
as in the Fierz $D,M$ in~\eqref{32} and~\eqref{33}. Following the notation in Fig. \ref{fig:Bocoord}
\begin{equation}
    \bm{r}_{c\bar{q}}=\bm{r}_{\bar{c}\bar{q}}+\bm{r}_{c\bar{c}}, \quad \bm{r}_{\bar{c}q}=\bm{r}_{cq}-\bm{r}_{c\bar{c}}\,.
\end{equation}

In the general case we want to study here, the coefficients $a_i(r_{c\bar{c}})$ are obtained through the variational principle by minimizing the energy of the $q\bar{q}$ system determined from the following hamiltonian:
\begin{equation}
    H_{q\bar q}=-\frac{\nabla_{\bm{r}_{cq}}^2}{2m_{cq}}-\frac{\nabla_{\bm{r}_{\bar{c}\bar{q}}}^2}{2m_{cq}}+V^{\text{coul}}+V^{\text{conf}}+V^{\text{spin}}\,.
    \label{eq:Hqq}
\end{equation}
We will discuss in detail the various terms appearing in this formula in Secs.~\ref{sec:coulomb}, \ref{sec:conf}, and~\ref{sec:spin}. Moreover we assume that the oribital angular momenta vanish so that $P=+1$ and $J=S$. 

The $a_i$ coefficients are obtained solving 
\be
\frac{d}{d a_i}\langle\Psi_{q\bar{q}}|H_{q\bar{q}}|\Psi_{q\bar{q}}\rangle=0
\label{eq:variational}
\ee
which can also be used to compute the Born-Oppenheimer potential $\Delta E(r_{c\bar{c}})$. It can be shown that  Eq.~\eqref{eq:variational} is equivalent to\footnote{Given a linear decomposition for $\Psi_{q\bar{q}}$ in terms of a set of functions $\{\Phi_i\}$ as
\be
    \Psi_{q\bar{q}}=\sum_{i=1}^N a_i\Phi_i
\ee
with $\sum_i |a_i|^2=1$, then \eqref{eq:variational} is equivalent to solving the problem
\be
    \sum_{j=1}^N\left(\langle\Phi_i|H_{q\bar{q}}|\Phi_j\rangle-\Delta E(r_{c\bar{c}})\langle\Phi_i|\Phi_j\rangle\right)a_j=0
\ee 
called the generalized eigenvalue problem \cite{eigenvalue}. If $\langle\Phi_i|\Phi_j\rangle=\delta_{ij}$, it reduces to the classic eigenvalue problem.}
\be
\begin{pmatrix}
        H_D-\Delta E(r_{c\bar{c}}) & H_{DM}-S_{DM}^2(r_{c\bar{c}})\,\Delta E(r_{c\bar{c}})\\
        H_{DM}-S_{DM}^2(r_{c\bar{c}})\,\Delta E(r_{c\bar{c}}) & H_M-\Delta E(r_{c\bar{c}})
    \end{pmatrix}
    \begin{pmatrix}
        a_1\\
        a_2
    \end{pmatrix}=0
\label{46}
\ee
where we introduced the overlap integral 
\begin{equation}
S_{DM}(r_{c\bar{c}})=\int_{\bm{r}_{cq}}\psi_D(\bm{r}_{cq})\psi_M(\bm{r}_{cq}-\bm{r}_{c\bar{c}})=\int_{\bm{r}_{\bar{c}\bar{q}}}\psi_D(\bm{r}_{\bar{c}\bar{q}})\psi_M(\bm{r}_{\bar{c}\bar{q}}-\bm{r}_{c\bar{c}})\,.
\end{equation}
The entries in the matrix equation~\eqref{46} are defined by
\begin{align}
& H_D=\langle\psi_D(\bm{r}_{cq})\psi_D(\bm{r}_{\bar{c}\bar{q}})|H_{q\bar{q}}|\psi_D(\bm{r}_{cq})\psi_D(\bm{r}_{\bar{c}\bar{q}})\rangle\nonumber\\
&H_M=\langle\psi_M(\bm{r}_{c\bar{q}})\psi_M(\bm{r}_{\bar{c}q})|H_{q\bar{q}}|\psi_M(\bm{r}_{c\bar{q}})\psi_M(\bm{r}_{\bar{c}q})\rangle\label{eq:H_matrix_element}\\
&H_{DM}=\langle\psi_D(\bm{r}_{cq})\psi_D(\bm{r}_{\bar{c}\bar{q}})|H_{q\bar{q}}|\psi_M(\bm{r}_{c\bar{q}})\psi_M(\bm{r}_{\bar{c}q})\rangle\nonumber
\end{align}
and their explicit expressions  are given in Sec.~\ref{sec:J_+1} and \ref{sec:J_-1}.

The BO potential is computed by requiring the vanishing of the determinant of the matrix~\eqref{46}, which ensures existence of a non-trivial solution (that is, different from $a_1=a_2=0$). This leads to
\begin{multline}
    \Delta E_{\pm}(r_{c\bar{c}})=\frac{1}{2(1-S_{DM}^4)}\Big(H_{D}+H_{M}-2S_{DM}^2\,H_{DM}\\\pm \sqrt{(H_D+H_{M}-2S_{DM}^2\,H_{DM})^2-4(1-S_{DM}^4)(H_{D}H_{M}-H_{DM}^2)}\:\Big)\,.
    \label{eq:lambda+-}
\end{multline}
For each of $\Delta E_-$ and $\Delta E_+$ one then obtains the relative size of  $a_1(r_{c\bar{c}})$ and $a_2(r_{c\bar{c}})$ from  the system of Eqs.~\eqref{46}. 

Next, before we go any further, we analyze in detail the expression for $H_{q\bar{q}}$ given by Eq.~\eqref{eq:Hqq} in order to provide the matrix elements of the system.

\subsection{Kinetic terms and Coulomb interactions}
\label{sec:coulomb}
We begin by analyzing the kinetic terms and the Coulomb interactions in the Hamiltonian $H_{q\bar{q}}$ in Eq.~\eqref{eq:Hqq}. The matrix elements arising from the kinetic term can be expressed in terms of the integral
\begin{equation}
    I_0^{DM}(R)=\int_{\bm{r}_{cq}}\psi_D(\bm{r}_{cq})(-\nabla^2_{\bm{r}_{cq}})\psi_M(\bm{r}_{cq}-\bm{r}_{c\bar{c}})=\int_{\bm{r}_{\bar{c}\bar{q}}}\psi_D(\bm{r}_{\bar{c}\bar{q}})(-\nabla^2_{\bm{r}_{\bar{c}\bar{q}}})\psi_M(\bm{r}_{\bar{c}\bar{q}}+\bm{r}_{c\bar{c}})
\end{equation}
and
\begin{equation}
    I_0^D=\lim_{M\to D}\lim_{r_{c\bar{c}}\to0}I_0^{DM}(r_{c\bar{c}})=D^2
\end{equation}
\begin{equation}
    I_0^M=\lim_{D\to M}\lim_{r_{c\bar{c}}\to0}I_0^{DM}(r_{c\bar{c}})=M^2
\end{equation}
The analytical expression for $I_0^{DM}$ is given in Appendix \ref{app:integrals}.

The Coulomb potential is obtained by treating the gluon dynamics in the one-gluon exchange approximation
\begin{equation}
    V^{\text{coul}}=-\frac{1}{3}\alpha_s\frac{1}{r_{cq}}-\frac{1}{3}\alpha_s\frac{1}{r_{\bar{c}\bar{q}}}-\frac{7}{6}\frac{1}{r_{c\bar{q}}}-\frac{7}{6}\alpha_s\frac{1}{r_{\bar{c}q}}+\frac{1}{6}\alpha_s\frac{1}{|\bm{r}_{cq} -\bm{r}_{\bar{c}\bar{q}}-\bm{r}_{c\bar{c}}|}
\end{equation}
The matrix elements of the Coulomb interaction are expressed in terms of the so-called two-center integrals $I_i(r_{c\bar{c}})$ \cite{bj}, listed in Appendix \ref{app:integrals} with their analytic expressions.

The contributions to the terms $H_D$, $H_M$, and $H_{DM}$ in Eq.~\eqref{eq:H_matrix_element} due to the kinetic term $K$ and Coulomb potentials are
\begin{flalign}
    &\langle \psi_{D}(\bm{r}_{cq})\psi_D(\bm{r}_{\bar{c}\bar{q}})|K +V^{\text{coul}}|\psi_{D}(\bm{r}_{cq})\psi_D(\bm{r}_{\bar{c}\bar{q}})\rangle&&\nonumber\\&=\dfrac{1}{m_{cq}}I_0^D+\alpha_s\left(-2\dfrac{1}{3}I_1^D(0)-2\dfrac{7}{6}I_1^D(r_{c\bar{c}})+\dfrac{1}{6}I_4^{D}(r_{c\bar{c}})\right)&&
    \label{eq:kinetic_A}
\end{flalign}
\begin{flalign}
    &\langle \psi_{M}(\bm{r}_{c\bar{q}})\psi_M(\bm{r}_{\bar{c}q})|K+V^{\text{coul}}|\psi_{M}(\bm{r}_{c\bar{q}})\psi_M(\bm{r}_{\bar{c}q})\rangle&&\nonumber\\&=\dfrac{1}{m_{cq}}I_0^M+\alpha_s\left(-2\dfrac{7}{6}I_1^M(0)-2\dfrac{1}{3}I_2^M(r_{c\bar{c}})+\dfrac{1}{6}I_4^{M}(r_{c\bar{c}})\right) &&
    \label{eq:kinetic_B}
\end{flalign}
\begin{flalign}
    &\langle \psi_{D}(\bm{r}_{cq})\psi_D(\bm{r}_{\bar{c}\bar{q}})|K +V^{\text{coul}}|\psi_{M}(\bm{r}_{c\bar{q}})\psi_M(\bm{r}_{\bar{c}q})\rangle=&&\nonumber\\&=S_{DM}(r_{c\bar{c}})\left(\dfrac{1}{m_{cq}}I_0^{DM}(r_{c\bar{c}})-2\dfrac{1}{3}\alpha_sI_2^{DM}(r_{c\bar{c}})-2\dfrac{7}{6}\alpha_sI_2^{MD}(r_{c\bar{c}})\right)+\frac{1}{6}\alpha_s I_6^{DM}(r_{c\bar{c}})&&
\end{flalign}
Note that the first two terms in each of~\eqref{eq:kinetic_A} and~\eqref{eq:kinetic_B} correspond precisely to the first two terms in $E_\mathcal{C}$, cf Eq.~\eqref{eq:E_C}.

\subsection{Confinement terms}
\label{sec:conf}
We do not have an explicit expression for this potential as in the case of  $V^{\text{coul}}$. Consider the matrix element $\langle\psi_D(\bm{r}_{cq})\psi_D(\bm{r}_{\bar{c}\bar{q}})|V^{\text{conf}}|\psi_D(\bm{r}_{cq})\psi_D(\bm{r}_{\bar{c}\bar{q}})\rangle$, which enters in the calculation of $H_D$ in \eqref{eq:H_matrix_element}. In configuration $D$, heavy and light (anti)quarks are arranged in (anti)diquarks. We will assume therefore that the previous matrix element will depend on the integral
\begin{equation}
    J^D_2=\int_{\bm{r}_{cq}}\psi_D^2(\bm{r}_{cq})\,r_{cq}=\int_{\bm{r}_{\bar{c}\bar{q}}}\psi_D^2(\bm{r}_{\bar{c}\bar{q}})\,r_{\bar{c}\bar{q}}.
\end{equation}
whereas $\langle\psi_M(\bm{r}_{c\bar{q}})\psi_M(\bm{r}_{\bar{c}q})|V^{\text{conf}}|\psi_M(\bm{r}_{c\bar{q}})\psi_M(\bm{r}_{\bar{c}q})\rangle$ in $H_M$ is associated with the ‘meson-antimeson’ configuration ($c\bar{q}$ and $\bar{c}q$) and it will depend on
\begin{equation}
    J^M_2=\int_{\bm{r}_{cq}}\psi_M^2(\bm{r}_{cq}-\bm{r}_{c\bar{c}})\,|\bm{r}_{cq}-\bm{r}_{c\bar{c}}|=\int_{\bm{r}_{\bar{c}\bar{q}}}\psi_M^2(\bm{r}_{\bar{c}\bar{q}}+\bm{r}_{c\bar{c}})\,|\bm{r}_{\bar{c}\bar{q}}+\bm{r}_{c\bar{c}}|.
\end{equation}
The tetraquark is  formed by a $c\bar cq\bar q$ cluster of quarks at small relative distances. Two color correlations are possible $[cq][\bar c\bar q]$ ($\bm 3,\bar{\bm 3}+\bm 6,\bar{\bm 6}$) or $(c\bar q)(\bar c q)$ ($\bm 1,\bm 1+\bm 8,\bm 8$). 
The energy needed to increase asymptotically the distance $r_{c\bar{c}}$ would create quark pairs from vacuum and lead to different configurations with respect to  $T_D$ or $T_M$, appearing in the two diagonal entries of~\eqref{46}. At $r_{c\bar{c}}$ of the order of the hadron size, the tetraquark is in the superposition  of Eq.~\eqref{41} and off-diagonal elements of~\eqref{46} have to be taken into account. Since the quark-antiquark color cofigurations have stronger attraction, we assume that the confinement potential in the off-diagonal terms scales as the $c\bar q$ quark-antiquark distance. We assume therefore that the off-diagonal matrix elements will be proportional to
\begin{equation}\label{242}
    J_2^{MD}(r_{c\bar{c}})=\int_{\bm{r}_{cq}}\psi_D(\bm{r}_{cq})\psi_M(\bm{r}_{cq}-\bm{r}_{c\bar{c}}) |\bm{r}_{cq}-\bm{r}_{c\bar{c}}|=\int_{\bm{r}_{\bar{c}\bar{q}}}\psi_D(\bm{r}_{cq})\psi_M(\bm{r}_{\bar{c}\bar{q}}+\bm{r}_{c\bar{c}}) |\bm{r}_{\bar{c}\bar{q}}+\bm{r}_{c\bar{c}}|.
\end{equation}
The analytic expression for the $J_2$ integrals is reported in Appendix \ref{app:integrals}.

As was done in the previous section, we write down the matrix elements explicitly
\begin{eqnarray}
    \langle\psi_D(\bm{r}_{cq})\psi_D(\bm{r}_{\bar{c}\bar{q}})|V^{\text{conf}}|\psi_D(\bm{r}_{cq})\psi_D(\bm{r}_{\bar{c}\bar{q}})\rangle&=&2\,k\,J_2^D\nonumber\\
    \langle\psi_M(\bm{r}_{c\bar{q}})\psi_M(\bm{r}_{\bar{c}q})|V^{\text{conf}}|\psi_M(\bm{r}_{c\bar{q}})\psi_M(\bm{r}_{\bar{c}q})\rangle&=&2\,k\,J_2^M\\
    \langle\psi_D(\bm{r}_{cq})\psi_D(\bm{r}_{\bar{c}\bar{q}})|V^{\text{conf}}|\psi_M(\bm{r}_{c\bar{q}})\psi_M(\bm{r}_{\bar{c}q})\rangle&=&2\,k\,S_{DM}(r_{c\bar{c}})\,J^{MD}_2(r_{c\bar{c}})\nonumber
\end{eqnarray}
The off-diagonal contributions are zero when $r_{c\bar{c}} \to\infty$, because the integral $J^{MD}(r_{c\bar{c}})$ vanishes.

\subsection{Spin interaction}
\label{sec:spin}
The quark potential used in \cite{x3872} does not account for the spin of the quarks, so it cannot distinguish between the $X(3872)$ and the other tetraquarks with different spin. In this work we add account for spin information through the potential \cite{isgur,cromomagn}
\be
    V^{\text{spin}}=\sum_{\text{pairs}}-\frac{\lambda_{ij}}{m_im_j}\frac{8\pi}{3}\alpha_s\,\delta^{(3)}\left(\boldsymbol{r}_i-\boldsymbol{r}_j\right)\,\boldsymbol{S}_i\cdot\boldsymbol{S}_j=\sum_{\text{pairs}}\mathcal{K}_{ij}(\textbf{R}^\textbf{c}_{ij})\,\boldsymbol{S}_i\cdot\boldsymbol{S}_j
    \label{eq:Vspintesto}
\ee
where $\textbf{R}^\textbf{c}_{ij}$ is the color representation of the $ij$ pair. The sum is extended over all pairs except for $c\bar{c}$.

In this article, we study the cases $J=0,1$, and $2$. We restrict our attention to states with zero orbital angular momentum ($L=0$) and therefore a state's total spin $J$ is precisely given by the quarks' total spin, $J=S$. Furthermore,  $P=+1$  for all states.  Denoting by $(c\bar c)^{\textbf{R}}_S$ a pair with total spin $S$ in the $\textbf{R}$ color representation, and by $|\psi\rangle^{\phantom{\dagger}}_J$ a state with total spin $J$,  tetraquark combinations with total spin $J=0$ can be written as $|(c\bar{c})^{\boldsymbol{8}}_0(q\bar{q})^{\boldsymbol{8}}_0\rangle^{\phantom{\dagger}}_0$ or $|(c\bar{c})^{\boldsymbol{8}}_1(q\bar{q})^{\boldsymbol{8}}_1\rangle^{\phantom{\dagger}}_0$, both with $C=+1$. 
To obtain  $J=1$,  possible combination are  $|(c\bar{c})^{\boldsymbol{8}}_0(q\bar{q})^{\boldsymbol{8}}_1\rangle^{\phantom{\dagger}}_1,|(c\bar{c})^{\boldsymbol{8}}_1(q\bar{q})^{\boldsymbol{8}}_0\rangle^{\phantom{\dagger}}_1,\,\text{and }|(c\bar{c})^{\boldsymbol{8}}_1(q\bar{q})^{\boldsymbol{8}}_1\rangle^{\phantom{\dagger}}_1$. The first two of these have $C=-1$, while the last one has $C=+1$. The case $J=2$ is obtained from $|(c\bar{c})^{\boldsymbol{8}}_1(q\bar{q})^{\boldsymbol{8}}_1\rangle^{\phantom{\dagger}}_2$, which has $C=+1$. 

To simplify the treatment, we first consider the color-spin part without the spatial wave functions. It can be shown (see Appendix~\ref{app:Spin}) that the effect due to the spin and color quantum numbers is to produce the following potentials
\begin{align}
    &V^{\text{spin},\pm}(0^{++})=\sum_{\text{pairs}}\;_0\langle(c\bar{c})^{\mathbf{8}}_1(q\bar{q})^{\mathbf{8}}_1|\mathcal{K}_{ij}(\textbf{R}^\textbf{c}_{ij})\,\boldsymbol{S}_i\cdot\boldsymbol{S}_j|(c\bar{c})^{\mathbf{8}}_1(q\bar{q})^{\mathbf{8}}_1\rangle^{\phantom{\dagger}}_0=\nonumber\\&=\frac{1}{16}\left(\pm2\sqrt{49\mathcal{K}_{\bar{c}q}^2(\bm{1})-28\mathcal{K}_{\bar{c}q}(\bm{1})\left(\mathcal{K}_{cq}(\bar{\bm{3}})+\mathcal{K}_{q\bar{q}}(\bm{8})\right)+16\left(\mathcal{K}_{cq}^2(\bar{\bm{3}})-\mathcal{K}_{cq}(\bar{\bm{3}})\mathcal{K}_{q\bar{q}}(\bm{8})+\mathcal{K}_{q\bar{q}}^2(\bm{8})\right)}\right.\nonumber\\&\,\,\quad\quad\quad-7\mathcal{K}_{\bar{c}q}(\bm{1})-4\big(\mathcal{K}_{cq}(\bar{\bm{3}})+\mathcal{K}_{q\bar{q}}(\bm{8})\big)\bigg) \label{eq:Spin0}\\
    &V^{\text{spin}}(1^{++})=-\frac{7}{16}\mathcal{K}_{\bar{c}q}(\mathbf{1})-\frac{1}{4}\mathcal{K}_{cq}(\mathbf{\bar{3}})+\frac{1}{4}\mathcal{K}_{q\bar{q}}(\mathbf{8})
    \label{eq:Vspinridotta}\\
    &V^{\text{spin},\pm}(1^{+-})=\pm\frac{1}{9}\sqrt{\left(\frac{63}{16}\mathcal{K}_{\bar{c}q}(\bm{1})-\frac{9}{4}\mathcal{K}_{cq}(\bar{\bm{3}})\right)^2+\frac{81}{4}\mathcal{K}_{q\bar{q}}^2(\bm{8})}-\frac{1}{4}\mathcal{K}_{q\bar{q}}(\bm{8})
    \label{eq:V_qq_neg}\\
    &V^{\text{spin}}(2^{++})=\frac{7}{16}\mathcal{K}_{\bar{c}q}(\bm{1})+\frac{1}{4}\big(\mathcal{K}_{cq}(\bar{\bm{3}})+\mathcal{K}_{q\bar{q}}(\bm{8})\big)
    \label{eq:Spin2}
\end{align}
where we have used the relations between the quadratic Casimirs of $\SU(3)$ to express the $\mathcal{K}_{ij}$ in terms of the representations $\bm{1},\,\bm{\bar{3}}$ and $\bm{8}$.

Now we introduce the spatial part. We define
\begin{eqnarray}
    K_{ij}(D;\textbf{R}_{ij}^\textbf{c})&=&\langle \psi_D(\bm{r}_{cq})\psi_D(\bm{r}_{\bar{c}\bar{q}})|\mathcal{K}_{ij}(\textbf{R}_{ij}^\textbf{c})|\psi_D(\bm{r}_{cq})\psi_D(\bm{r}_{\bar{c}\bar{q}})\rangle\nonumber\\
    K_{ij}(M;\textbf{R}_{ij}^\textbf{c})&=&\langle \psi_M(\bm{r}_{c\bar{q}})\psi_M(\bm{r}_{\bar{c}q})|\mathcal{K}_{ij}(\textbf{R}_{ij}^\textbf{c})|\psi_M(\bm{r}_{c\bar{q}})\psi_M(\bm{r}_{\bar{c}q})\rangle\\
    K_{ij}(D,M;\textbf{R}_{ij}^\textbf{c})&=&\langle \psi_D(\bm{r}_{cq})\psi_D(\bm{r}_{\bar{c}\bar{q}})|\mathcal{K}_{ij}(\textbf{R}_{ij}^\textbf{c})|\psi_M(\bm{r}_{c\bar{q}})\psi_M(\bm{r}_{\bar{c}q})\rangle\nonumber
\end{eqnarray}
All these involve matrix elements of a potential with the Dirac $\delta$ function, and therefore most, but not all,  can be  written simply in terms of wavefunctions at a single point. For example
\begin{multline}
    K_{c\bar{q}}(D,{\boldsymbol{1}})=\langle\psi_D(\bm{r}_{cq})\psi_D(\bm{r}_{\bar{c}\bar{q}})|\mathcal{K}_{c\bar{q}}(\mathbf{1})|\psi_D(\bm{r}_{cq})\psi_D(\bm{r}_{\bar{c}\bar{q}})\rangle\\\propto\langle\psi_D(\bm{r}_{cq})\psi_D(\bm{r}_{\bar{c}\bar{q}})|\delta(\bm{r}_{cq})|\psi_D(\bm{r}_{cq})\psi_D(\bm{r}_{\bar{c}\bar{q}})\rangle=\psi_D^2(0)
\end{multline}
The explicit form of these $K$'s is reported in Appendix~\ref{app:K}. 

The potentials to be used in $H_D$, $H_M$, and $H_{DM}$ are obtained by substituting $\mathcal{K}_{ij}$ in \eqref{eq:Vspinridotta} and \eqref{eq:V_qq_neg} with $K_{ij}(D;\textbf{R}_{ij}^\textbf{c})$, $K_{ij}(M;\textbf{R}_{ij}^\textbf{c})$, and $K_{ij}(D,M;\textbf{R}_{ij}^\textbf{c})$, respectively.

In the next sections, we will explicitly compute the various spin contributions for the spectrum of the $c\bar{c}q\bar{q}$ tetraquarks.

\subsection{The tetraquark spectrum}
The spin contributions split the degeneracy  of the spectrum. In listing the quantum numbers one should bear in mind that $P=+1$ and $J=S$, since orbital angular momenta vanish.

We start with the $1^{++}$ state because, as we will see, it can be associated with the $X(3872)$ particle, which we take as a reference point for calculating the other masses.  
\subsubsection{\texorpdfstring{$J^{PC}=1^{++}$}{J=1+}}
\label{sec:J_+1}
The spin interactions for the $1^{++}$ case reduce to
\begin{flalign}
    V_{D,1^{++}}^{\text{spin}}&=\langle\psi_D(\bm{r}_{cq})\psi_D(\bm{r}_{\bar{c}\bar{q}})|V^{\text{spin}}(1^{++})|\psi_D(\bm{r}_{cq})\psi_D(\bm{r}_{\bar{c}\bar{q}})\rangle=    \nonumber\\
    &=-\frac{7}{16}K_{c\bar{q}}(D;\boldsymbol{1}) - \frac{1}{4}K_{cq}(D;\bar{\boldsymbol{3}})+\frac{1}{4}K_{q\bar{q}}(D;\boldsymbol{8})\nonumber\\
    V_{M,1^{++}}^{\text{spin}}&=\langle\psi_M(\bm{r}_{c\bar{q}})\psi_M(\bm{r}_{\bar{c}q})|V^{\text{spin}}(1^{++})|\psi_M(\bm{r}_{c\bar{q}})\psi_M(\bm{r}_{\bar{c}q})\rangle=\label{eq:Vbspin}\\
    &=-\frac{7}{16}K_{c\bar{q}}(M;\boldsymbol{1}) - \frac{1}{4}K_{cq}(M;\bar{\boldsymbol{3}}) + \frac{1}{4}K_{q\bar{q}}(M;\boldsymbol{8})\nonumber\\
    V_{DM,1^{++}}^{\text{spin}}&=\langle\psi_D(\bm{r}_{cq})\psi_D(\bm{r}_{\bar{c}\bar{q}})|V^{\text{spin}}(1^{++})|\psi_M(\bm{r}_{c\bar{q}})\psi_M(\bm{r}_{\bar{c}q})\rangle= \nonumber\\
    &=-\frac{7}{16}K_{c\bar{q}}(D,M;\boldsymbol{1}) - \frac{1}{4}K_{cq}(D,M;\bar{\boldsymbol{3}}) + \frac{1}{4}K_{q\bar{q}}(D,M;\boldsymbol{8})\nonumber
\end{flalign}
and the matrix elements are
\begin{align}
& H_{D,1^{++}}=2  E_D  - 2\frac{7}{6}\alpha_s I_2^D(r_{c\bar{c}}) + \frac{1}{6}\alpha_s\,I_4^D(r_{c\bar{c}}) + V_{D,1^{++}}^{\text{spin}}(r_{c\bar{c}})\nonumber\\
&H_{M,1^{++}}=2 E_M  - 2\frac{1}{3}\alpha_s I_2^M(r_{c\bar{c}}) + \frac{1}{6}\alpha_s\,I_4^D(r_{c\bar{c}}) + V_{M,1^{++}}^{\text{spin}}(r_{c\bar{c}})\label{eq:matrix_elements}\\
&H_{DM,1^{++}}=S_{DM}(r_{c\bar{c}})\left(\dfrac{1}{m_{cq}} I_0^{DM}(r_{c\bar{c}}) - 2 \dfrac{1}{3} \alpha_s I_2^{DM}(r_{c\bar{c}}) - 2 \dfrac{7}{6} \alpha_s I_2^{MD}(r_{c\bar{c}}) + 2\, k\, J_2^{MD}(r_{c\bar{c}})\right) \nonumber\\
&\quad\quad\,\,\,\,\,\quad\quad + \frac{1}{6} \alpha_s\,I_6^{DM}(r_{c\bar{c}}) + V_{DM,1^{++}}^{\text{spin}}(r_{c\bar{c}})\nonumber
\end{align}
where $E_\mathcal{C}$ was defined in \eqref{eq:E_C}.
Using these in~\eqref{eq:lambda+-} we obtain the potentials $\Delta E^\pm_{1^{++}}(r_{c\bar{c}})$ to be included in the equation of the charm quarks.

The coefficients $a_{i,1^{++}}^\pm(r_{c\bar{c}})$ in Eq.~\eqref{41} still need to be calculated in order to determine the eigenfunctions of the light quarks. We find it more convenient, in general for all spins, to switch to an orthonormal basis, 
\begin{align}
    \Psi_{q\bar{q},1} &= \frac{\psi_D(\bm{r}_{cq})\psi_D(\bm{r}_{\bar{c}\bar{q}}) + \psi_M(\bm{r}_{c\bar{q}})\psi_M(\bm{r}_{\bar{c}q})}{\sqrt{2(1+S^2_{DM}(r_{c\bar{c}}))}} \\
    \Psi_{q\bar{q},2} &=  \frac{\psi_D(\bm{r}_{cq})\psi_D(\bm{r}_{\bar{c}\bar{q}}) - \psi_M(\bm{r}_{c\bar{q}})\psi_M(\bm{r}_{\bar{c}q})}{\sqrt{2(1-S^2_{DM}(r_{c\bar{c}}))}}
\end{align}
such that $\Psi_{q\bar{q}}$ is written as
\begin{equation}
    \Psi_{q\bar{q}} = \frac1{\sqrt{c_1^2(r_{c\bar{c}})+c_2^2(r_{c\bar{c}})}}\bigg(c_1(r_{c\bar{c}})  \Psi_{q\bar{q},1}  +  c_2(r_{c\bar{c}})\Psi_{q\bar{q},2} \bigg)
\end{equation}
and the problem reduces to a classic eigenvalue problem.

\begin{figure}
    \centering
    \includegraphics[width=0.5\linewidth]{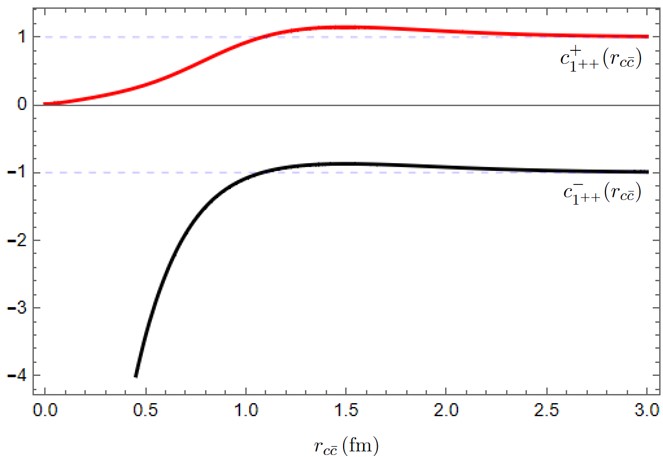}
    \caption{Behavior of the ratios $c^\pm(r_{c\bar{c}})$ defined in \eqref{eq:c1_c2p} as a function of the distance $r_{c\bar{c}}$ between the charm quarks for the case $J^{PC}=1^{++}$. The dashed horizontal lines highlight the limits to which the two functions asymptotically converge.}
    \label{fig:c1+-}
\end{figure}
In this new basis, the equation \eqref{46} for the coefficients $c_i(r_{c\bar{c}})$ becomes, independently of the spin\footnote{The eigenvalues $\Delta E^\pm$ are obviously independent of the chosen basis. This fact can be verified by calculating $\Delta E^\pm$ from \eqref{eq:matrix_epsilon} and comparing the result with \eqref{eq:lambda+-}.}
\be
    \begin{pmatrix}
        \epsilon_{11}(r_{c\bar{c}})-\Delta E^\pm(r_{c\bar{c}}) & -\epsilon_{12}(r_{c\bar{c}})\\
        -\epsilon_{12}(r_{c\bar{c}}) & \epsilon_{22}(r_{c\bar{c}})-\Delta E^\pm(r_{c\bar{c}})
    \end{pmatrix}
    \begin{pmatrix}
        c_1^\pm\\
        c_2^\pm
    \end{pmatrix}=0
\label{eq:matrix_epsilon}    
\ee 
where
\begin{eqnarray}
    \epsilon_{11}(r_{c\bar{c}})&=&\frac{H_{M}(r_{c\bar{c}})+H_{D}(r_{c\bar{c}})+2H_{DM}(r_{c\bar{c}})}{2+2[S^2_{DM}(r_{c\bar{c}})]}\\
    \epsilon_{22}(r_{c\bar{c}})&=&\frac{H_{M}(r_{c\bar{c}})+H_{D}(r_{c\bar{c}})-2H_{DM}(r_{c\bar{c}})}{2-2[S^2_{DM}(r_{c\bar{c}})]}\\
    \epsilon_{12}(r_{c\bar{c}})&=&\frac{H_{D}(r_{c\bar{c}})-H_{M}(r_{c\bar{c}})}{2\sqrt{1-S^4_{DM}(r_{c\bar{c}})}}
\end{eqnarray}
The expressions for the ratios $c_1^\pm(r_{c\bar{c}})/c_2^\pm(r_{c\bar{c}})$ corresponding to the two eigenvalues $\Delta E^\pm(r_{c\bar{c}})$ are
\begin{equation}
    c^\pm(r_{c\bar{c}})=\frac{c^\pm_1(r_{c\bar{c}})}{c^\pm_2(r_{c\bar{c}})}=\frac{\epsilon_{22}-\epsilon_{11}}{2\epsilon_{12}}\pm\sqrt{1+\left(\frac{\epsilon_{22}-\epsilon_{11}}{2\epsilon_{12}}\right)^2}
    \label{eq:c1_c2p}
\end{equation}
which allows to define the normalized wavefunction for the light quarks as
\begin{equation}
    \Psi^{\pm}_{q\bar{q}}=\frac{1}{\sqrt{1+(c^{\pm})^2}}\bigg(c^{\pm}(r_{c\bar{c}})  \Psi_{q\bar{q},1}+\Psi_{q\bar{q},2} \bigg)
    \label{eq:psipm}
\end{equation}
 
\begin{figure}[t!]
    \centering
    \includegraphics[width=0.55\textwidth]{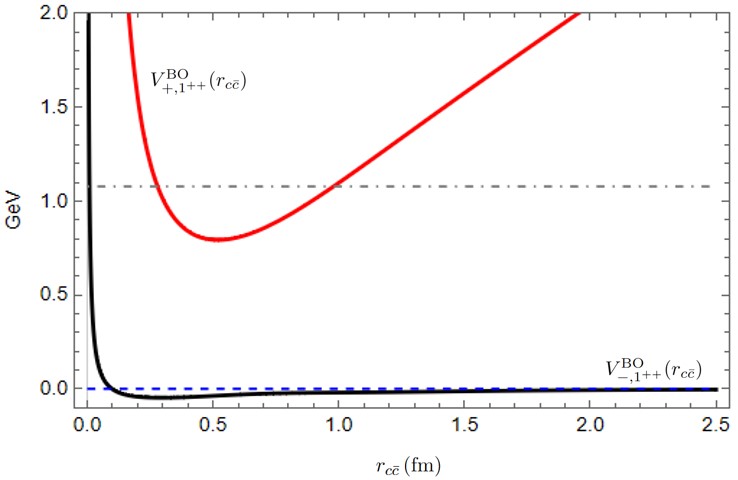}
    \caption{The  potentials $V_{\pm,1^{++}}^{\rm{BO}}$ described by Eqs.~\eqref{eq:VBO-} and \eqref{eq:VBO+} for the $c\bar{c}$ pair, along with the energies of their respective ground states (dashed and dot-dashed lines). The potentials are vertically shifted so that $V_{-,1^{++}}^{\rm{BO}}(r_{c\bar{c}})\to0$ asymptotically. We associate the ground state of the $V_{-,1^{++}}^{\rm{BO}}(r_{c\bar{c}})$ potential, with an energy of approximately $-3$ MeV, with the $X(3872)$ particle, while the other state is roughly $1.1$ GeV higher in energy (corresponding to a mass around $5.0$ GeV).}
    \label{fig:vbocc}
\end{figure}

The behavior of the ratios $c^\pm_{1^{++}}$ for the case $J^{PC}=1^{++}$ is shown in Fig.~\ref{fig:c1+-} and determines the behavior of the wavefunctions $\Psi^{\pm}_{q\bar{q},1^{++}}$ which, in turn, as we will see shortly, determine the confinement potential between the $c\bar{c}$ quarks:
\begin{itemize}
    \item For $r_{c\bar{c}}\to+\infty$ the ratio $c^-_{1^{++}}$ goes to $-1$. This means that for a large separation of the heavy quarks, $\Psi^-_{q\bar{q},C=+1}$ tends to
    \begin{align}
       & -\frac{\psi_D(\bm{r}_{cq})\psi_D(\bm{r}_{\bar{c}\bar{q}})+\psi_M(\bm{r}_{c\bar{q}})\psi_M(\bm{r}_{\bar{c}q})}{\sqrt{2(1+S_{DM}(+\infty)^2)}}+\frac{\psi_D(\bm{r}_{cq})\psi_D(\bm{r}_{\bar{c}\bar{q}})-\psi_M(\bm{r}_{c\bar{q}})\psi_M(\bm{r}_{\bar{c}q})}{\sqrt{2(1-S_{DM}(+\infty)^2)}}\nonumber\\&\propto \psi_M(\bm{r}_{c\bar{q}})\psi_M(\bm{r}_{\bar{c}q})
    \end{align}
   We can interpret this result as saying that the tetraquark tends to separate in such a way that a meson-meson pair is formed. However we recall that the $\psi_M$ orbital is not what we would have if the $c-\bar{q}$ system were in a color singlet because it accounts also for the octet interactions (it is calculated using $\lambda_{c\bar{q}}=-7/6$, see Eq.~\eqref{eq:casimir} and \eqref{eq:lambda_cbarq}). Still, if we consider a tensor $T^i_j$ in the \textbf{8} of $\SU(3)_C$ this can be contracted with a tensor of a soft gluon $A^j_i$ and produce a singlet. For large distances, the \textbf{8} component is screened by soft gluons and therefore we expect that no confining potential arises. 
   \item By contrast, for $r_{c\bar{c}}\to+\infty$ the ratio $c^+_{1^{++}}$ goes to $+1$ and so $\Psi_{q\bar{q},1^{++}}^+$ goes to
    \begin{align}
        &\frac{\psi_D(\bm{r}_{cq})\psi_D(\bm{r}_{\bar{c}\bar{q}})+\psi_M(\bm{r}_{c\bar{q}})\psi_M(\bm{r}_{\bar{c}q})}{\sqrt{2(1+S_{DM}(+\infty)^2)}}+\frac{\psi_D(\bm{r}_{cq})\psi_D(\bm{r}_{\bar{c}\bar{q}})-\psi_M(\bm{r}_{c\bar{q}})\psi_M(\bm{r}_{\bar{c}q})}{\sqrt{2(1-S_{DM}(+\infty)^2)}}\nonumber\\&\propto \psi_D(\bm{r}_{cq})\psi_D(\bm{r}_{\bar{c}\bar{q}})
    \end{align}
   In this case, the tetraquark separates into a diquark-antidiquark pair; therefore, we have to include a confining potential to ensure color neutrality.
\end{itemize}

In light of these, we have the following two potentials for the $c\bar{c}$ pairs:
\begin{equation}
    V_{-,1^{++}}^{\rm{BO}}(r_{c\bar{c}})=\frac{1}{6}\alpha_s\frac{1}{r_{c\bar{c}}}+\Delta E^-_{1^{++}}(r_{c\bar{c}})
    \label{eq:VBO-}
\end{equation}
\begin{equation}
    V_{+,1^{++}}^{\rm{BO}}(r_{c\bar{c}})=\frac{1}{6}\alpha_s\frac{1}{r_{c\bar{c}}}+\Delta E^+_{1^{++}}(r_{c\bar{c}})+k \,r_{c\bar{c}}
    \label{eq:VBO+}
\end{equation}
which are shown in Fig.~\ref{fig:vbocc} (in the figure, the potentials are shifted so that $V_{-,1^{++}}^{\rm{BO}}$ tends to 0 asymptotically). The $V_-^{\rm{BO}}$ case corresponds to the $(\bm 1,\bm 1+{\bm 8,\bm 8})$ asymptotic configuration, whereas the $V_+^{\rm{BO}}$ case is related to the asymptotic case $(\bm 3, \bar{\bm{3}}+\bm 6,\bar{\bm 6})$.

Both potentials admit a bound state whose energies are shown in Fig.~\ref{fig:vbocc} together with the respective potentials. The lowest state, which is located at an energy of $\simeq -3$ MeV, can be identified with the $X(3872)$. The upper one is approximately $1.1$ GeV above it which means that its mass is around $5.0$ GeV, which makes it highly unstable. The non-relativistic approximation may be questionable in this case.
Our calculated mass splitting $m(X(3872))-m(D\bar D^*)=-3$\,MeV deviates from precise measurements~\cite{Tomaradze:2015cza,pdg}; a similar result is found in \cite{nora}. Despite this discrepancy, the resulting value remains remarkably small considering that the typical energy scales involved in the problem, such as the separation between the lower and upper ground states, are much larger, and that the model has not been fine tuned to obtain the small mass splitting.

We note two differences with respect to~\cite{x3872}. The only asymptotic configuration assumed in~\cite{x3872}  was: $\bm{3},\bar{\bm{3}}+\bm 6,\bar{\bm 6}$ (see Sec.~\ref{sec:BOT}, Eq.~\eqref{41}). This would correspond to associating the $X(3872)$ with the ground state of the potential $V_{+,1^{++}}^{\text{BO}}$, instead of $V_{-,1^{++}}^{\text{BO}}$. Furthermore, in Ref.~\cite{x3872} the linearly  rising confining term for the $c\bar{c}$ pair has an onset at some $r_{c\bar{c}} \geq R_{0s}= 3\pm1\,\text{fm}$ --- in contrast, in this paper $R_{0s}=0$.\footnote{This implies that we can not recover  the same wavefunction as in \cite{x3872}, even if we turn off the spin potentials and set $a_2(r_{c\bar{c}})=0$ in Eq.~\eqref{41}.} We remark that $R_{0s}$ changes the offset of $V_+^{\rm BO}$ only: a value of $R_{0s}\simeq 3$~fm would give a $\approx 200$~MeV mass gap between the bound state in $V_-^{\rm BO}$ and that in $V_+^{\rm BO}$ whereas $R_{0s}=0$ allows a larger mass gap of $\approx 1$~GeV. Given the absence of narrow charmed (non-strange) exotic $1^{++}$ states within $\approx 700$~MeV above the $X$ mass, we are led to assume $R_{0s}=0$.

\begin{figure}[t!]
        \centering
        \subfloat{\includegraphics[width=0.42\textwidth]{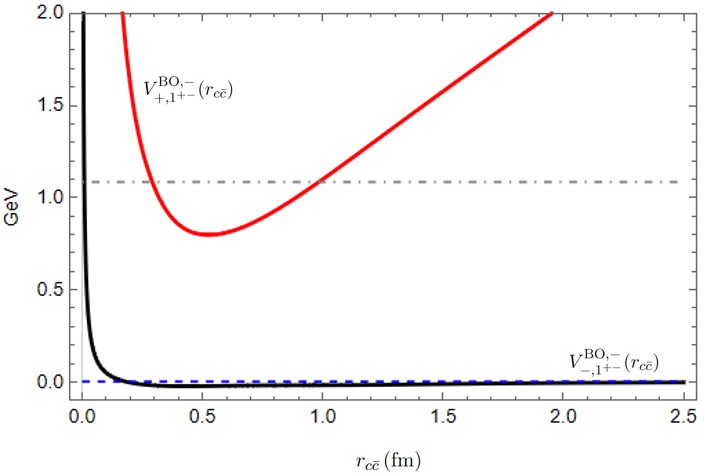}}
        \qquad
        \subfloat{\includegraphics[width=0.42\textwidth]{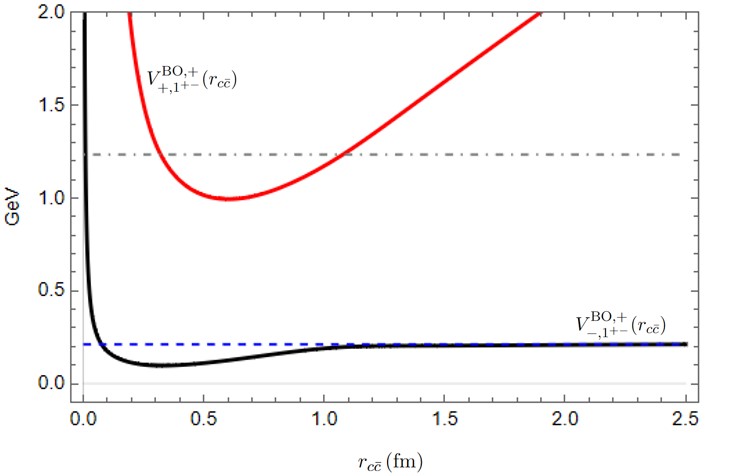}}\\
        \caption{The  potentials $V_{\pm,1^{+-}}^{\rm{BO},\pm}$ described by Eqs.~\eqref{eq:V_BO_spin-} and \eqref{eq:V_BO_spin+} for the $c\bar{c}$ pair, along with the energy of their respective ground states (dashed and dot-dashed lines). The potentials have the same shift as those in Fig.~\ref{fig:vbocc}, so that the energies can be compared directly. We associate the lowest state of the left potential with the particle $Z_c(3900)$, while for the potential on the right we refer to Sec.~\ref{sez:summary}, where two possible assignments are indicated.}
        \label{fig:Z_Zp}
\end{figure}

\subsubsection{\texorpdfstring{$J^{PC}=1^{+-}$}{J=1-}}
\label{sec:J_-1}
The spin interactions for the $1^{+-}$ case produce two different potentials (see Eq.~\eqref{eq:V_qq_neg}):
\begin{flalign}
    V_{D,1^{+-}}^{\text{spin},\pm}&=\langle\psi_D(\bm{r}_{cq})\psi_D(\bm{r}_{\bar{c}\bar{q}})|V^{\text{spin},\pm}(1^{+-})|\psi_D(\bm{r}_{cq})\psi_D(\bm{r}_{\bar{c}\bar{q}})\rangle= \nonumber\\
    &=\pm\frac{1}{9}\sqrt{\left(\frac{63}{16}K_{\bar{c}q}(D;\bm{1})-\frac{9}{4}K_{cq}(D;\bm{3})\right)^2+\frac{81}{4}K_{q\bar{q}}^2(D;\bm{8})}-\frac{1}{4}K_{q\bar{q}}(D;\bm{8})\nonumber\\
    V_{M,1^{+-}}^{\text{spin},\pm}&=\langle\psi_M(\bm{r}_{c\bar{q}})\psi_M(\bm{r}_{\bar{c}q})|V^{\text{spin},\pm}(1^{+-})|\psi_M(\bm{r}_{c\bar{q}})\psi_M(\bm{r}_{\bar{c}q})\rangle= \nonumber\\
    &=\pm\frac{1}{9}\sqrt{\left(\frac{63}{16}K_{\bar{c}q}(M;\bm{1})-\frac{9}{4}K_{cq}(M;\bm{3})\right)^2+\frac{81}{4}K_{q\bar{q}}^2(M;\bm{8})}-\frac{1}{4}K_{q\bar{q}}(M;\bm{8})\\
    V_{DM,1^{+-}}^{\text{spin},\pm}&=\langle\psi_D(\bm{r}_{cq})\psi_D(\bm{r}_{\bar{c}\bar{q}})|V^{\text{spin},\pm}(1^{+-})|\psi_M(\bm{r}_{c\bar{q}})\psi_M(\bm{r}_{\bar{c}q})\rangle= \nonumber\\
    &=\pm\frac{1}{9}\sqrt{\left(\frac{63}{16}K_{\bar{c}q}(D,M;\bm{1})-\frac{9}{4}K_{cq}(D,M;\bm{3})\right)^2+\frac{81}{4}K_{q\bar{q}}^2(D,M;\bm{8})}-\frac{1}{4}K_{q\bar{q}}(D,M;\bm{8})\nonumber
\end{flalign}
The matrix elements can be calculated from Eq.~\eqref{eq:matrix_elements} in the previous section by substituting the spin potentials with those written here.

The same considerations made for $J^C=1^+$ also apply here. Therefore, the four (two for each spin potential) heavy-quarks potentials are
\begin{equation}
    V_{-,1^{+-}}^{\rm{BO},\pm}(r_{c\bar{c}})=\frac{1}{6}\alpha_s\frac{1}{r_{c\bar{c}}}+\Delta E^{-,\pm}_{1^{+-}}(r_{c\bar{c}})
    \label{eq:V_BO_spin-}
\end{equation}
\begin{equation}
    V_{+,1^{+-}}^{\rm{BO},\pm}(r_{c\bar{c}})=\frac{1}{6}\alpha_s\frac{1}{r_{c\bar{c}}}+\Delta E^{+,\pm}_{1^{+-}}(r_{c\bar{c}})+k \,r_{c\bar{c}}
    \label{eq:V_BO_spin+}
\end{equation}
The upper index $\pm$ depends on the spin potential used. The $V_-^{\rm{BO},\pm}$ cases once again correspond to the $(\bm 1,\bm 1+{\bm 8,\bm 8})$ asymptotic configuration, and the $V_+^{\rm{BO},\pm}$ cases are related to the asymptotic case $(\bm 3, \bar{\bm 3}+\bm 6,\bar{\bm 6})$. The four potentials are shown in Fig.~\ref{fig:Z_Zp} along with the energy of the ground states (the same vertical shift applied in Fig.~\ref{fig:vbocc} is also used in these figures).

As in the previous section, the particles associated to the $V_{+,1^{+-}}^{\rm{BO}}$ potentials are $O(\text{GeV})$ heavier than the corresponding particles associated to $V_{-,1^{+-}}^{\rm{BO}}$, making them highly unstable.

The ground state of the potential $V_{-,1^{+-}}^{\rm{BO},-}(r_{c\bar{c}})$ is nearly degenerate with the one of the potential $V_{-,1^{++}}^{\rm{BO}}(r_{c\bar{c}})$ of the previous section (cf. Fig.~\ref{fig:vbocc} and \ref{fig:Z_Zp}), which we  associated with the $X(3872)$. Therefore, we identify this state with the $Z_c(3900)$, which has the appropriate quantum numbers.

For the ground state of the potential $V_{-,1^{+-}}^{\rm{BO},+}(r_{c\bar{c}})$, we propose two possible assignments, which are discussed in Sec.~\ref{sez:summary}.

\subsubsection{\texorpdfstring{$J^{PC}=0^{++}$}{J=0++}}
\begin{figure}[t!]
        \centering
        \subfloat{\includegraphics[width=0.42\textwidth]{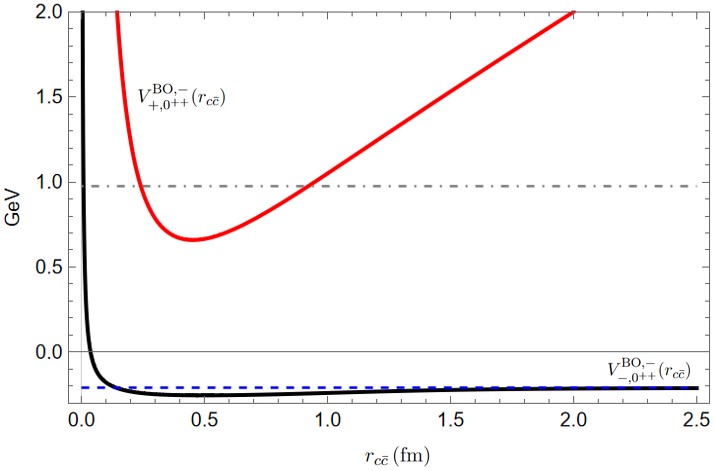}}
        \qquad
        \subfloat{\includegraphics[width=0.42\textwidth]{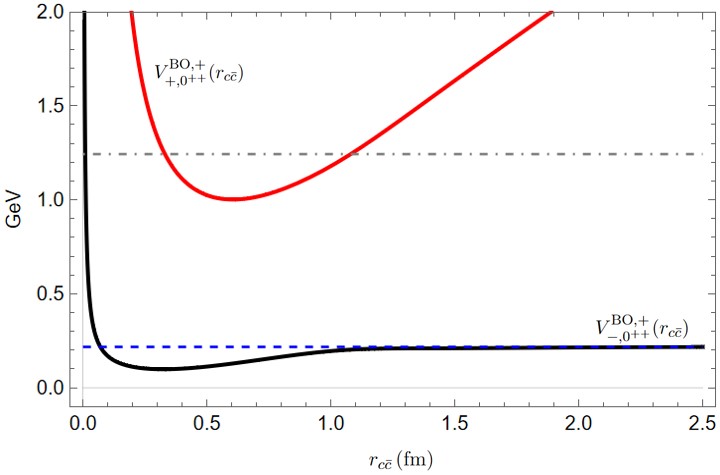}}\\
        \caption{The potentials $V_{\pm,0^{++}}^{\rm{BO},\pm}$ described by Eqs.~\eqref{eq:V_BO_spin0-} and \eqref{eq:V_BO_spin0+} for the $c\bar{c}$ pair, along with the energy of their respective ground states (dashed and dot-dashed lines). The potentials have the same shift as those in Fig.~\ref{fig:vbocc}, so that the energies can be compared directly. We associate the lowest state of the right potential with the particle $X(4100)$, while for the left lower state there are currently no tetraquark candidates with the correct quantum numbers and within the corresponding mass range.}
        \label{fig:spin0}
\end{figure}
Much as in the $1^{+-}$ case, the $0^{++}$ also has two different spin potentials, which are obtained from the potentials \eqref{eq:Spin0} by the substitution $\mathcal{K}_{ij}\to K_{ij}$, as seen in the previous sections.

The same considerations made for $1^{++}$  apply here. The four heavy-quarkss potentials are
\begin{equation}
    V_{-,0^{++}}^{\rm{BO},\pm}(r_{c\bar{c}})=\frac{1}{6}\alpha_s\frac{1}{r_{c\bar{c}}}+\Delta E^{-,\pm}_{0^{++}}(r_{c\bar{c}})
    \label{eq:V_BO_spin0-}
\end{equation}
\begin{equation}
    V_{+,0^{++}}^{\rm{BO},\pm}(r_{c\bar{c}})=\frac{1}{6}\alpha_s\frac{1}{r_{c\bar{c}}}+\Delta E^{+,\pm}_{0^{++}}(r_{c\bar{c}})+k \,r_{c\bar{c}}
    \label{eq:V_BO_spin0+}
\end{equation}
The upper index $\pm$ is related to the choice of $V^{\text{spin},\pm}(0^{++})$. The four potentials are shown in Fig.~\ref{fig:spin0} along with the energy of the ground states (the potentials have been shifted by the same amount as those in Fig.~\ref{fig:vbocc}).

The particles associated to the $V_{+,0^{++}}^{\rm{BO}}$ potentials are $O(\text{GeV})$ heavier than the corresponding particles associated to $V_{-,0^{++}}^{\rm{BO}}$, making them highly unstable. We do not have an assignment for the ground state of the potential $V_{-,0^{++}}^{\rm{BO},-}(r_{c\bar{c}})$ because approximately $215$ MeV below the $X(3872)$, there are currently no tetraquark candidates with the correct quantum numbers. On the other hand, the ground state of the potential $V_{-,0^{++}}^{\rm{BO},+}(r_{c\bar{c}})$, which is $\simeq210$ MeV above the $X(3872)$, can be associated with the $X(4100)$, which, as reported by the PDG, is compatible with the $0^{++}$ assignment.

\subsubsection{\texorpdfstring{$J^{PC}=2^{++}$}{J=2++}}
\label{sec:spin2}
\begin{figure}[t!]
    \centering
    \includegraphics[width=0.55\textwidth]{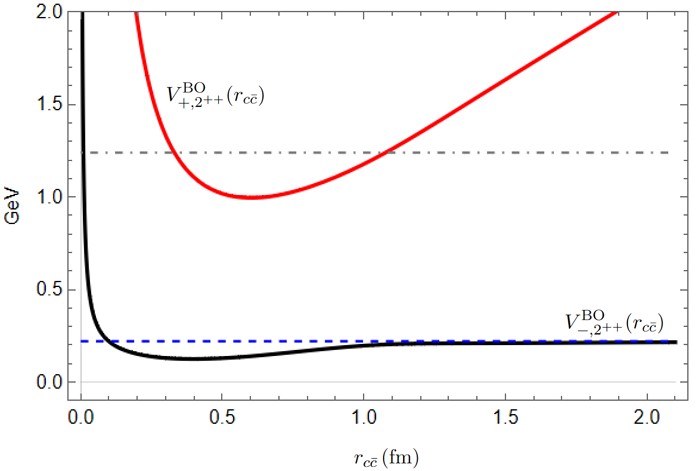}
    \caption{The  potentials $V_{\pm,2^{++}}^{\rm{BO}}$ described by Eqs.~\eqref{eq:V_BO_spin2-} and \eqref{eq:V_BO_spin2+} for the $c\bar{c}$ pair, along with the energies of their respective ground states (dashed and dot-dashed lines). The potentials have the same shift as those in Fig.~\ref{fig:vbocc}, so that the energies can be compared directly.}
    \label{fig:spin2}
\end{figure}
The considerations made in the previous sections also apply to the case $2^{++}$, for which we have a unique spin potential (just as in the $1^{++}$ case) given by \eqref{eq:Spin2} with the usual substitution $\mathcal{K}_{ij}\to K_{ij}$. The charm-anticharm potentials are
\begin{equation}
    V_{-,2^{++}}^{\rm{BO}}(r_{c\bar{c}})=\frac{1}{6}\alpha_s\frac{1}{r_{c\bar{c}}}+\Delta E^{-}_{2^{++}}(r_{c\bar{c}})
    \label{eq:V_BO_spin2-}
\end{equation}
\begin{equation}
    V_{+,2^{++}}^{\rm{BO}}(r_{c\bar{c}})=\frac{1}{6}\alpha_s\frac{1}{r_{c\bar{c}}}+\Delta E^{+}_{2^{++}}(r_{c\bar{c}})+k \,r_{c\bar{c}}
    \label{eq:V_BO_spin2+}
\end{equation}
which are shown in Fig.~\ref{fig:spin2} along with the energy of the ground states (with the same constant shift in potentials as applied in Fig.~\ref{fig:vbocc}).

The ground state of the potential $V_{-,2^{++}}^{\rm{BO}}$ is about $210$ MeV above the $1^{++}$. There are currently no tetraquark candidates with the correct quantum numbers in this mass range.  

In the next section, we summarize what has been discussed so far.

\subsubsection{Comparison of the spectrum to data}
\label{sez:summary}
\begin{figure}
    \centering
    \includegraphics[width=0.55\linewidth]{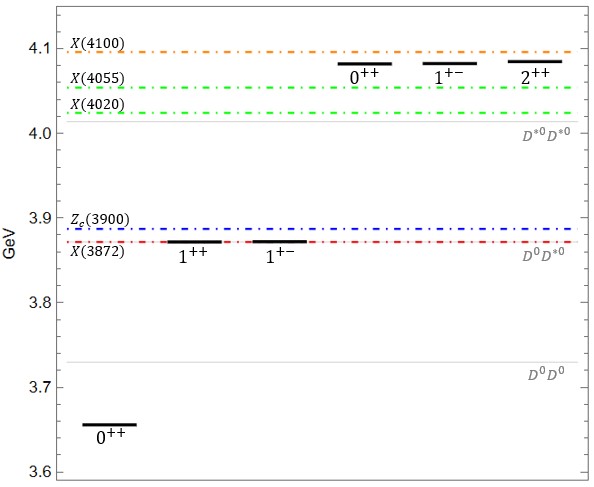}
    \caption{Tetraquark spectrum obtained in this article. Below each state, the quantum numbers $J^{PC}$. The solid lines represent the $D^0D^0$, $D^0D^{*0}$, and $D^{*0}D^{*0}$ thresholds, while the dot-dashed lines correspond to the experimental masses of the particles which are assigned to the states.}
    \label{fig:spectro}
\end{figure}

The whole spectrum determined by $V_{-}^{\rm{BO}}$ is shown in Fig.~\ref{fig:spectro}. In the same plot, the $D^0D^0$, $D^0D^{*0}$ and $D^{*0}D^{*0}$ thresholds and the experimental masses of the resonances relevant to our comparison  are included.

Our  $J^{PC}=1^{++}$ state is naturally associated  with the $X(3872)$, and  the $1^{+-}$, almost degenerate with the $1^{++}$, can be associated with the $Z_c(3900)$. The other $1^{+-}$ state, at about $210$ MeV above the $1^{++}$, can be associated either with the $X(4020)$ or with the $X(4050)$ with the caveat that the quantum numbers for these two resonances are still unknown. The experimental values for the mass differences are
\begin{eqnarray}
    M_{X(4020)}-M_{X(3872)}&=& 153\pm2\,\text{MeV}\\
    M_{X(4050)}-M_{X(3872)}&=&179^{+24}_{-40}\,\text{MeV} 
\end{eqnarray}

The heavier scalar could be the $X(4100)$, compatible with the $0^{++}$ assignment, according to the PDG.

As for the lightest scalar $0^{++}$ and the $2^{++}$ state, no exotic resonances compatible with our mass predictions have been observed so far.  

The quark model provides complete $I=0$  and $I=1$ isospin multiplets. The $X$ particle's proximity to the $D^0\bar D^{0*}$ threshold amplifies the effect of isospin breaking and therefore $X(3872)$ is likely  a very uneven superposition of the $X_u=c\bar c u\bar u$ and $X_d=c\bar c d\bar d$ quantum states, representing a strong  admixture of $I=0$ and $I=1$. One of these neutral states may fall below the di-meson threshold, making it observable, while the other, situated above it, may simply be too broad to be detected \cite{compact1}. Being a thorough admixture of $I=0$ and $I=1$, the lighter, visible one is expected to decay into $J/\psi\,\rho$ and $J/\psi\, \omega$.
The expected $X^\pm$ particles filling the $I=1$ multiplet, may fall below the $D^\pm \bar D^{*0}$ threshold forcing them to decay into $J/\psi \rho^\pm$. This may be difficult to detect because of the slow  $\pi^0$ \bg from the $\rho^\pm$ in the final states. On the other hand the charged $1^{+-}$ lower states, corresponding to $Z_c^{\pm}$, can decay into the more accessible $J/\psi\pi^{\pm}$ modes, as observed. As for the $Z_c^0$, also in this case only one state is found. We remind the reader that exotic hadrons are typically experimentally found to be in isospin triplets \cite{pdg}.

\section{Radiative decays of the \texorpdfstring{$X(3872)$}{X(3872)} revisited}
\label{sec:rad}
In~\cite{x3872} a calculation of
\be
    \mathcal{R}=\frac{\text{Br}(X\to\psi^\prime\,\gamma)}{\text{Br}(X\to\psi\,\gamma)}
\ee
for both the molecular and the compact tetraquark interpretations is proposed,  assuming a non-relativistic setting. 

\begin{figure}[t]
    \centering
    \includegraphics[width=0.36\textwidth]{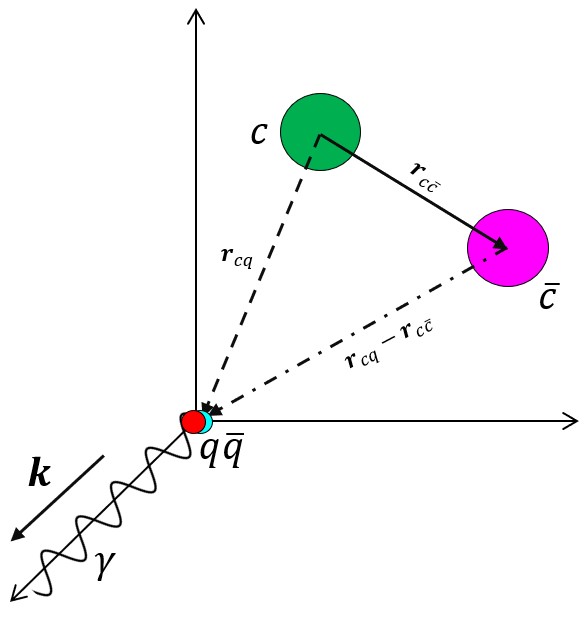}
    \caption{Scheme of the dominant process for the radiative decay $X\to\psi^{(\prime)}\gamma$. Light quarks annihilate at the origin producing a photon with momentum $\bm k$.}
    \label{fig:annhil}
\end{figure}

At the lowest order, the process $X\to\psi^{(\prime)}\gamma$ is dominated by the annihilation of the $q\bar{q}$ pair. Without loss of generality, we assume that the annihilation takes place in the origin of the frame in  Fig.~\ref{fig:annhil}. Defining $\psi(\bm{r}_{c\bar{c}})$ as the wave function of the final charmonium, which is a function of the distance $r_{c\bar{c}}$ between the $c\bar{c}$ pair,  the transition amplitude $A$ in the $X$ rest frame, at a fixed photon three-momentum $\bm k$,  is given by\footnote{Only the real part of the exponential factor contributes to the amplitude \cite{x3872}
\begin{equation}
    A\left(X\to\psi^{(\prime)}\gamma\right)=\mathcal{F}\int_{\bm{r}_{c\bar{c}},\bm{r}_{cq}} \cos\left[k\left(\cos\eta\left(\frac{R}{2}-\xi\cos\theta\right)-\xi\sin\theta\sin\eta\cos\phi\right)\right]
    \psi_c(r_{c\bar{c}})\Psi_{c\bar{c}q\bar{q}}
(\bm{r}_{c\bar{c}},\bm{r}_{cq})\notag
\end{equation}
choosing the frame orientations such that: $\bm{r}_{cq}=(\xi\sin\theta\cos\phi,\xi\sin\theta\sin\phi,\xi\cos\theta),\,\bm{r}_{c\bar{c}}=(0,0,R)$ and $\bm{k}=(k\sin\eta,0,k\cos\eta)$}
\begin{equation}  
A\left( X\to\psi^{(\prime)}\gamma\right)=
\mathcal{F}\int_{\bm{r}_{c\bar{c}},\bm{r}_{cq}} e^{-i\bm k\cdot\left(\frac{\bm{r}_{c\bar{c}}}{2}-\bm{r}_{cq}\right)}\,\psi(|\bm{r}_{c\bar{c}}|)\Psi_{c\bar{c}q\bar{q}}
(\bm{r}_{c\bar{c}},\bm{r}_{cq})
\label{eq:A}
\end{equation}
where both in the molecular and compact models one assumes the $X$ wavefunction is factorized, $\Psi_{c\bar{c}q\bar{q}}(\bm{r}_{c\bar{c}},\bm{r}_{cq})=\Psi_{c\bar{c}}(|\bm{r}_{c\bar{c}}|)\Psi_{q\bar{q}}(|\bm{r}_{cq}|,|\bm{r}_{cq}-\bm{r}_{c\bar{c}}|)$.
The factor $\mathcal{F}$ takes into account those common factors which cancel out in the calculation of the ratio ${\mathcal R}$.
The value of $\mathcal{R}$ depends only on the ratio of the squared moduli of the amplitudes \eqref{eq:A}, the ratio of phase spaces $\Phi=0.26$, and the sum over polarizations $\mathcal{P}=0.98$, which do not cancel out in the ratio since they  depend on the momentum ${\bm k}$ of the produced photon, 
\be
    |\bm k|=\frac{M_X^2-M^2_{\psi^{(\prime)}}}{2M_X}\,.
\ee
Putting everything together yields
\be
\mathcal{R}=\Phi\,\mathcal{P}\left|\frac{A\left(X\to\psi^{\prime}\gamma\right)}{A\left(X\to\psi\gamma\right)}\right|^2
\approx0.25\left|\frac{A\left(X\to\psi^{\prime}\gamma\right)}{A\left(X\to\psi\gamma\right)}\right|^2
\label{eq:Rteorica}
\ee

\begin{figure}[t]
        \centering
        \includegraphics[width=0.55\textwidth]{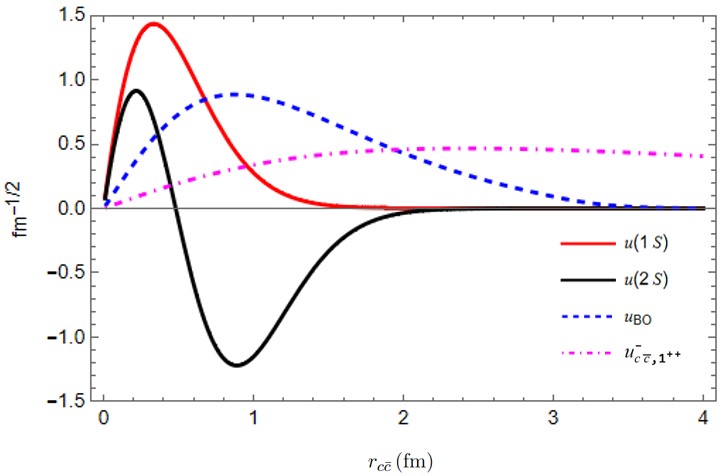}
        \caption{Normalized reduced radial wave functions for the charmonia $\psi(1S)$ and $\psi(2S)$ (or $\psi^\prime$) and for the $c\bar{c}$ pair: $u_{\text{BO}}$ is the wavefunction obtained in \cite{x3872} considering only $\psi_D$ orbitals; $u_{c\bar{c},1^{++}}^-$ is the wavefunction computed in this article using the potential $V_{-,1^{++}}^{\rm{BO}}$. The function $u_{c\bar{c},1^{++}}^-$ presents a peak shifted more to the right and a longer tail compared to $u_{\text{BO}}$.}
        \label{fig:funzradiali}
\end{figure}

In~\cite{x3872} the wavefunction $\Psi_{q\bar{q}}$ for a compact tetraquark is assumed to be
\begin{equation}
\Psi_{q\bar{q}}(\bm{r}_{cq},\bm{r}_{\bar{c}\bar{q}},\bm{r}_{c\bar{c}})=\psi_D(\bm{r}_{cq})\psi_D(\bm{r}_{\bar{c}\bar{q}})
\label{ilcompatto}
\end{equation}
because only the diquark-antidiquark case was considered, and consequently $\Psi_{c\bar{c}}(\bm{r}_{c\bar{c}})$ was calculated in this approximation.
In this paper we have considered also the $\psi_M$ orbitals (meson-antimeson configuration). In the previous sections, we concluded that the $X(3872)$ can be associated with the ground state of the potential $V_{-,1^{++}}^{\rm{BO}}(r_{c\bar{c}})$ in Eq.~\eqref{eq:VBO-}, thus $\Psi_{c\bar{c}}$ is obtained by solving the Schrödinger equation with this potential (which we define $\Psi_{c\bar{c},1^{++}}^-$), and  is given, in our variational solution,  by (cfr. Eq.~\eqref{eq:psipm})
\begin{multline}
    \Psi^{-}_{q\bar{q},1^{++}}=\frac{1}{\sqrt{1+(c^{-}_{1^{++}})^2}}\left(c^{-}_{1^{++}}(r_{c\bar{c}})\frac{\psi_D(\bm{r}_{cq})\psi_D(\bm{r}_{\bar{c}\bar{q}})+\psi_M(\bm{r}_{c\bar{q}})\psi_M(\bm{r}_{\bar{c}q})}{\sqrt{2(1+S^2_{DM}(r_{c\bar{c}}))}}\right.\\\left.+\frac{\psi_D(\bm{r}_{cq})\psi_D(\bm{r}_{\bar{c}\bar{q}})-\psi_M(\bm{r}_{c\bar{q}})\psi_M(\bm{r}_{\bar{c}q})}{\sqrt{2(1-S^2_{DM}(r_{c\bar{c}}))}}\right)
\end{multline}
In Fig.~\ref{fig:funzradiali}, we have compared the reduced radial wavefunctions for the $c\bar{c}$ pair: $u_{c\bar{c},1^{++}}^-$ is calculated in this paper, while $u_{\text{BO}}$ was obtained in \cite{x3872}. On the same plot, we have shown the wavefunctions for the 1S and 2S charmonia. The function $u_{c\bar{c},1^{++}}^-$ peaks at higher values of the distance $r_{c\bar{c}}$ and has a longer tail compared to $u_{\text{BO}}$.

Using the functions $\Psi_{q\bar{q},1^{++}}^-$ and $\Psi_{c\bar{c},1^{++}}^-$ in Eq.~\eqref{eq:Rteorica}, we obtain the following estimate for $\mathcal{R}$:
\begin{equation}
    \mathcal{R}^-_{1^{++}} = 1.4 \pm 0.3
\end{equation}
The associated uncertainty is obtained by varying $m_q$ and $k$ (only in \eqref{eq:Hqq}) in Eq.~\eqref{eq:parametri} by $\pm10\%$.\footnote{These two parameters determine the size of the orbitals of the light quarks and, consequently, the size of the tetraquark. In \cite{x3872}, it was shown that the size of the tetraquark plays a fundamental role in the ratio $\mathcal{R}$.}
This value ought to be compared with the latest measurement by LHCb \cite{LHCb}, which reports
\be
    \mathcal{R}_{\text{exp}}=1.67\pm0.21\pm0.12\pm0.04.
\ee

\section{Summary}
\label{sez:concl}

We have updated and improved the discussion in \cite{x3872} on the compact tetraquark hypothesis. This  requires introducing a configuration consisting of a pair of colored meson-like states  in addition to the diquark-antidiquark configuration, as outlined in Eq.~\eqref{41}. This extension leads to the derivation of two distinct potentials for the heavy quarks: one corresponding to the colored mesonlike pair configuration, which is energetically favorable, and another corresponding to the  diquark pair configuration. Furthermore, we incorporate the spin interaction \eqref{eq:Vspintesto}, enabling us to distinguish states based on their total spin and charge conjugation. Within this framework, we constructed all possible $J^{PC}$ combinations with vanishing orbital angular momentum: $0^{++},\,1^{++},\,1^{+-}$, and $2^{++}$. We identified the $1^{++}$ state with the $X(3872)$ --- as summarized in Sec.~\ref{sez:summary}, where we also provide assignments for the other states.

Recently, two alternative approaches to the calculation of the $c\bar{c}q\bar{q}$ tetraquark spectrum have been proposed in \cite{nora,lattice_Braaten}. Both employ the Born-Oppenheimer approximation, but the heavy quark potential is obtained using the static energies of the light quarks derived from lattice calculations \cite{static_energy1,static_energy2,static_energy3}, in contrast to our methodology. 

A direct comparison with our spectrum is not feasible, as \cite{nora,lattice_Braaten} focus on $c\bar{c}q\bar{q}$ tetraquarks with $I=0$, while our method does not allow for the distinction of isospin. Nonetheless, all three studies predict a $0^{++}$ state below the $X(3872)$ and a $2^{++}$ state near the $D^{*0}D^{*0}$ threshold, neither of which has been experimentally observed so far. In our work, the mass of the $0^{++}$ state is approximately $3.65 \, \text{GeV}$, whereas in \cite{nora,lattice_Braaten}, it is around $3.8 \, \text{GeV}$. The mass of the $2^{++}$ state is around $4.08\,\text{GeV}$ compared to $4.01\,\text{GeV}$ in \cite{nora,lattice_Braaten}.

We obtain two $1^{+-}$ states, as in \cite{lattice_Braaten}, whereas \cite{nora} predicts only one. The lower one is (almost) degenerate with the $X(3872)$, so we identify it with the $Z_c(3900)$. This spectrum closely matches the experimental data.
The $J=0,1,2$ lower spectrum of tetraquarks found is composed of shallow bound states, which are linear superpositions of color $(\bm 1,\bm 1)+(\bm8,\bm 8)$ di-mesonlike states with a size closer to the typical hadron size. These compact shallow bound states are distinct from the molecular ones expected in non-relativistic scattering theory, as their fundamental features are determined by color forces and spin-interactions and their wavefunctions do not match the universal form predicted for a shallow molecular bound state \cite{lqm3-2}.

The $1^{++}$ and the $1^{+-}$ states closely reproduce the observed spectrum. The two $0^{++}$ states have a significantly larger mass gap than predicted in~\cite{compact1}.

By identifying the $1^{++}$ state with the $X(3872)$, we refined the calculation of $\mathcal{R} = \text{Br}(X \to \psi^\prime \gamma) / \text{Br}(X \to J/\psi \gamma)$ using the model proposed in \cite{x3872}. Our estimate for $\mathcal{R}$ (determined with the BO potential $V_{-,1^{++}}(r_{c\bar{c}})$) is (cf. Sec.~\ref{sec:rad})
\begin{equation}
    \mathcal{R}^{-}_{1^{++}} = 1.4 \pm 0.3
\end{equation}
which is consistent with the most recent measurement by the LHCb collaboration \cite{LHCb},
\begin{equation}
    \mathcal{R}_{\text{exp}} = 1.67 \pm 0.21 \pm 0.12 \pm 0.04.
\end{equation}
This result aligns with the findings in \cite{nora} as well.

In contrast, the molecular hypothesis discussed in \cite{molecradiative1,x3872} predicts $\mathcal{R}_{\text{mol}} \ll 1$ \cite{x3872}, which is clearly in contradiction with $\mathcal{R}_{\text{exp}}$.
It has also been pointed out in \cite{againstcc} that the radiative decays of the $X(3872)$ disfavor its interpretation as a pure $c\bar{c}$ state.

\section*{Acknowledgements}
We thank L. Maiani and V. Belayev for several discussions on this topic. The
work of B.G. is supported by the U.S. Department of Energy under
grant number DE-SC0009919. B.G. thanks the Instituto de Astrof\'isicas de Canarias and the Instituto de F\'isica Te\'orica, Madrid, where part of his work was done, for their hospitality. 

\newpage

\appendix

\section{\texorpdfstring{$I(R),S(R)$ and $J(R)$ functions}{I(R), S(R) and J(R) functions}}
\label{app:integrals}
\begin{figure}[ht]
    \centering
    \includegraphics[width=0.45\textwidth]{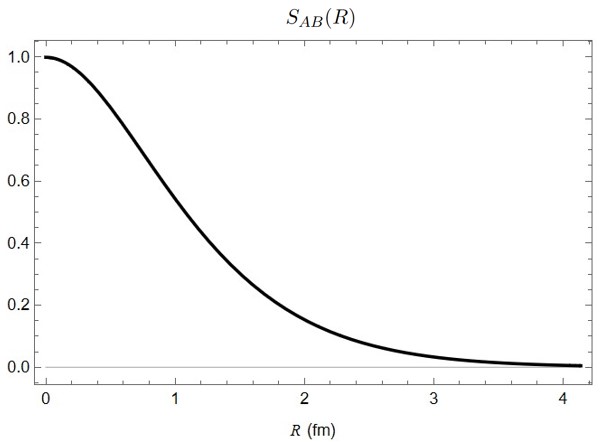}
    \caption{Behavior of the overlap integral $S_{AB}(R)$ \eqref{eq:overlap} for $A=D$ and $B=M$ defined in Eqs. \eqref{eq:E_A} and \eqref{eq:E_B}.}
    \label{fig:SAB}
\end{figure}
The function $\psi_{\mathcal{C}}(\bm{\zeta})$ is defined as (Eq. \eqref{varfunc})
\begin{equation}
    \psi_{\mathcal{C}}(\bm{\zeta})=\sqrt{\frac{\mathcal{C}^3}{\pi}}\,e^{-\mathcal{C}\,r}
\end{equation}
Hereafter $A,B>0$ to ensure the convergence of the integrals.

\textit{\textbf{Overlap integral:}}
\begin{align}
    S_{AB}(R)&=\int_{\bm{\xi}}\psi_A(\bm{\xi})\psi_B(\bm{\xi}-\bm{R})=\int_{\bm{\eta}}\psi_A(\bm{\eta})\psi_B(\bm{\eta}+\bm{R})=\nonumber\\&=\frac{8\sqrt{A^3B^3}}{R(A^2-B^2)^2}\left(R\left(B e^{-A R}+A e^{-B R}\right)+\frac{4 A B}{A^2-B^2}\left(e^{-AR}-e^{-BR}\right)\right)
    \label{eq:overlap}
\end{align}

\textit{\textbf{Kinetic term:}}
\begin{align}
    &I_0^{AB}(R)=\int_{\bm{\xi}}\psi_A(\bm{\xi})(-\nabla^2_{\bm{\xi}})\psi_B(\bm{\xi}-\bm{R})=\int_{\bm{\eta}}\psi_A(\bm{\eta})(-\nabla^2_{\bm{\eta}})\psi_B(\bm{\eta}+\bm{R})=\nonumber\\
    &=\frac{8 (AB)^{\frac{5}{2}}}{(A^2-B^2)^3} \left(e^{-B R} \left(\frac{2A^2+2B^2}{R}-A^2B+B^3\right)-e^{-AR}\left(\frac{2A^2+2B^2}{R}-AB^2+A^3\right)\right)
\end{align}
From this integral, we derive the integrals $I_0^A$
\begin{equation}
\label{app:I0A}
    I_0^{A}=\int_{\bm{\xi}}\psi_{A}(\bm{\xi})(-\nabla^2_{\bm{\xi}})\psi_{A}(\bm{\xi})=\int_{\bm{\eta}}\psi_{A}(\bm{\eta})(-\nabla^2_{\bm{\eta}})\psi_A(\bm{\eta})=\lim_{B\to A}\lim_{R\to0}I_0^{AB}(R)=A^2
\end{equation}
and $I_0^B$
\begin{equation}
    I_0^B=\int_{\bm{\xi}}\psi_{B}(\bm{\xi}-\bm{R})(-\nabla^2_{\bm{\xi}})\psi_{B}(\bm{\xi}-\bm{R})=\int_{\bm{\eta}}\psi_{B}(\bm{\eta}+\bm{R})(-\nabla^2_{\bm{\eta}})\psi_B(\bm{\eta}+\bm{R})=B^2
\end{equation}

\begin{figure}[t]
    \centering
    \includegraphics[width=0.5\textwidth]{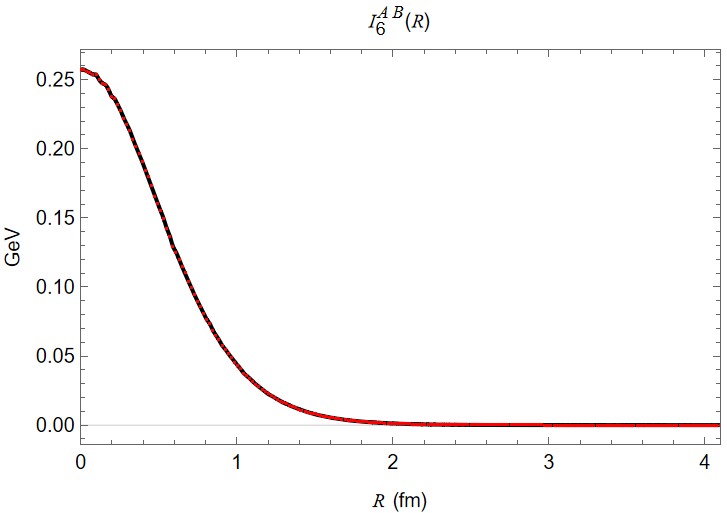}
    \caption{The red dots represent the result of the numerical integration of Eq.~\eqref{eq:I6abapp} performed with \texttt{Wolfram Mathematica} varying $R$ in the interval $[0,20]\,\text{fm}$ in steps of $0.02\,\text{fm}$ for $A=D$ and $B=M$ defined in Eqs. \eqref{eq:E_A} and \eqref{eq:E_B}. The black line shows the behavior of the interpolating function.}
    \label{fig:I6AB}
\end{figure}

\textit{\textbf{Two-center integrals:}}
The two-center integrals are known analytically \cite{bj}. $I_1^A(R)$ and $I_4^A(R)$ are
\begin{flalign}
\label{app:I1A}    I_1^A(R)=\int_{\boldsymbol{\xi}}\psi_A^2(\boldsymbol{\xi})\frac{1}{|\boldsymbol{\xi}-\boldsymbol{R}|}=\int_{\boldsymbol{\eta}}\psi_A^2(\boldsymbol{\eta})\frac{1}{|\boldsymbol{\eta}+\boldsymbol{R}|}=\frac{1}{R}-A\,e^{-2AR}\left(1+\frac{1}{AR}\right)
\end{flalign}
\begin{equation}
    I_4^A(R)=\int_{\boldsymbol{\xi},\boldsymbol{\eta}}\psi_A^2(\boldsymbol{\xi})\psi_A^2(\boldsymbol{\eta})\frac{1}{|\boldsymbol{\xi}-\boldsymbol{R}-\boldsymbol{\eta}|}=A\left(\frac{1}{AR}-e^{-2AR}\left(\frac{1}{AR}+\frac{11}{8}+\frac{3}{4}AR+\frac{1}{6}A^2R^2\right)\right)
\end{equation}
The integrals $I_i^B(R)$
\begin{eqnarray}
    I_1^B(R)&=&\int_{\boldsymbol{\xi}}\psi_B^2(\boldsymbol{\xi}-\boldsymbol{R})\frac{1}{\xi}=\int_{\boldsymbol{\xi}^\prime}\psi_B^2(\boldsymbol{\xi}^\prime)\frac{1}{|\boldsymbol{\xi}^\prime+\boldsymbol{R}|}=\int_{\bar{\boldsymbol{\xi}}}\psi_B^2(\bar{\boldsymbol{\xi}})\frac{1}{|\bar{\boldsymbol{\xi}}-\boldsymbol{R}|}\\
    I_4^B(R)&=&\int_{\boldsymbol{\xi},\boldsymbol{\eta}}\psi_B^2(\boldsymbol{\xi}-\boldsymbol{R})\psi_B^2(\boldsymbol{\eta}+\boldsymbol{R})\frac{1}{|\boldsymbol{\xi}-\boldsymbol{R}-\boldsymbol{\eta}|}=\int_{\boldsymbol{\xi}^\prime,\boldsymbol{\eta}^\prime}\psi_B^2(\boldsymbol{\xi}^\prime)\psi_B^2(\boldsymbol{\eta}^\prime)\frac{1}{|\boldsymbol{\xi}^\prime-\boldsymbol{R}-\boldsymbol{\eta}^\prime|}\hspace{1.3cm}
\end{eqnarray}
can be recast in the same form as the previous ones with a change of variables
\begin{eqnarray}
    I_1^B(R)&=&\int_{\boldsymbol{\xi}}\psi_B^2(\boldsymbol{\xi}-\boldsymbol{R})\frac{1}{\xi}=\int_{\bar{\boldsymbol{\xi}}}\psi_B^2(\bar{\boldsymbol{\xi}})\frac{1}{|\bar{\boldsymbol{\xi}}+\boldsymbol{R}|}=\int_{\boldsymbol{\xi}^\prime}\psi_B^2(\boldsymbol{\xi}^\prime)\frac{1}{|\boldsymbol{\xi}^\prime-\boldsymbol{R}|}\\
    I_4^B(R)&=&\int_{\boldsymbol{\xi},\boldsymbol{\eta}}\psi_B^2(\boldsymbol{\xi}-\boldsymbol{R})\psi_B^2(\boldsymbol{\eta}+\boldsymbol{R})\frac{1}{|\boldsymbol{\xi}-\boldsymbol{R}-\boldsymbol{\eta}|}=\int_{\boldsymbol{\xi}^\prime,\boldsymbol{\eta}^\prime}\psi_B^2(\boldsymbol{\xi}^\prime)\psi_B^2(\boldsymbol{\eta}^\prime)\frac{1}{|\boldsymbol{\xi}^\prime-\boldsymbol{R}-\boldsymbol{\eta}^\prime|}\hspace{1.3cm}
\end{eqnarray}
The integral $I_2^{AB}(R)$ is
\begin{align}
    I_2^{AB}(R)&=\int_{\boldsymbol{\xi}}\psi_A(\boldsymbol{\xi})\psi_B(\boldsymbol{\xi}-\boldsymbol{R})\frac{1}{|\boldsymbol{\xi}|}=\int_{\boldsymbol{\eta}}\psi_A(\boldsymbol{\eta})\psi_B(\boldsymbol{\eta}+\boldsymbol{R})\frac{1}{|\boldsymbol{\eta}|}=\nonumber\\
   &= 4\frac{\sqrt{A^3B^3}}{R}\left(\frac{R}{A^2-B^2}e^{-B R}+\frac{2 B}{(A^2-B^2)^2}\left(e^{-A R}-e^{-B R}\right)\right)
\end{align}
An analytical expression for $I_6^{AB}(R)$
\begin{equation}
    I_6^{AB}(R)=\int_{\boldsymbol{\xi},\boldsymbol{\eta}}\psi_A(\boldsymbol{\xi})\psi_A(\boldsymbol{\eta})\frac{1}{|\boldsymbol{\xi}-\boldsymbol{\eta}-\boldsymbol{R}|}\psi_B(\boldsymbol{\xi}-\boldsymbol{R})\psi_B(\boldsymbol{\eta}+\boldsymbol{R})
    \label{eq:I6abapp}
\end{equation}
is known only for $A=B$ \cite{I6}. Therefore, we estimated it numerically using \texttt{Wolfram Mathematica}\footnote{\texttt{NIntegrate} with the option \texttt{Method -> AdaptiveQuasiMonteCarlo}}, varying $R$ in the interval $[0,20]\,\text{fm}$ in steps of $0.02\,\text{fm}$. The result for $A=D$ and $B=M$ defined in Eqs. \eqref{eq:E_A} and \eqref{eq:E_B} is shown in Fig.~\ref{fig:I6AB} overlaid with the interpolating function obtained with \texttt{Interpolation}.

\textit{\textbf{Confinement integrals:}}
\begin{align}
    J_2^{BA}(R)&=\int_{\bm{\xi}}\psi_A(\bm{\xi})\psi_B(\bm{\xi-R}) |\bm{\xi}-\bm{R}|=\int_{\bm{\eta}}\psi_A(\bm{\eta})\psi_B(\bm{\eta+R}) |\bm{\eta}+\bm{R}|=\nonumber\\&=\frac{8(AB)^{3/2}}{(A^2-B^2)^4R}\bigg(Ae^{-BR}\left(\frac{4A^2+20B^2}{R}-8A^2B+8B^3+R(A^2-B^2)^2\right)+\nonumber\\ &\qquad\qquad\qquad\quad\,\,-e^{-AR}\left(\frac{4A^3-20AB^2}{R}+A^4+2A^2B^2-3B^4\right)\bigg)
\end{align}
From this integral, we derive
\begin{equation}
\label{app:J2A}
    J_2^A=\int_{\bm{\xi}}\psi_A^2(\bm{\xi}) \,\xi=\int_{\bm{\eta}}\psi_A^2(\bm{\eta})\,\eta=\lim_{B\to A}\lim_{R\to 0}J_2^{BA}(R)=\frac{2}{3 A}
\end{equation}
and
\begin{equation}
    J_2^{B}=\int_{\bm{\xi}}\psi_B^2(\bm{\xi}-\bm{R})|\bm{\xi}-\bm{R}|=\int_{\bm{\eta}}\psi_B^2(\bm{\eta})\, |\bm{\eta}+\bm{R}|
\end{equation}
which can be recast in the same form as the previous one with a change of variables
\begin{equation}
    J_2^{B}=\int_{\bm{\xi}}\psi_B^2(\bm{\xi}-\bm{R})\, |\bm{\xi}-\bm{R}|=\int_{\bm{\xi}^\prime}\psi_B^2(\bm{\xi}^\prime)\,\xi^\prime=\lim_{A\to B}\lim_{R\to0}J_2^{AB}(R)=\frac{2}{3B}
\end{equation}
\section{Spin Interactions}
\label{app:Spin}
In this Appendix we want to prove the Eqs.~\eqref{eq:Spin0}-\eqref{eq:Spin2} which provide the spin corrections in a $c\bar{c}q\bar{q}$ system due to interactions between light and heavy quarks (excluding the $c\bar{c}$ interaction which is extremely suppressed).

The spin interaction potential can be written in the following useful way
\begin{equation}
    V^{\text{spin}}=2\kappa^{\textbf{R}^\textbf{c}}_{cq}\,\mathbf{S}_c\cdot \mathbf{S}_q+2\kappa^{\textbf{R}^\textbf{c}}_{\bar{c}\bar{q}}\,\mathbf{S}_{\bar{c}}\cdot\mathbf{S}_{\bar{q}}+2\kappa^{\textbf{R}^\textbf{c}}_{q\bar{q}}\mathbf{S}_q\cdot\mathbf{S}_{\bar{q}}+2\kappa^{\textbf{R}^\textbf{c}}_{\bar{c}q}\mathbf{S}_{\bar{c}}\cdot\mathbf{S}_{q}+2\kappa^{\textbf{R}^\textbf{c}}_{c\bar{q}}\mathbf{S}_{c}\cdot\mathbf{S}_{\bar{q}}
    \label{eq:Vspin}
\end{equation}
where $2$ is included for convenience to eliminate the $1/2$ factor when transitioning from $\mathbf{S}_i\cdot\mathbf{S}_j$ to $\mathbf{S}^{tot}_{ij}$, and $\kappa^{\textbf{R}^\textbf{c}}_{ij}$ are the chromomagnetic factors for the $ij$ pair in the representation $\textbf{R}^\textbf{c}$. We assume the interactions to be charge conjugation invariant, so $\kappa^{\textbf{R}^\textbf{c}}_{Qq}=\kappa^{\bar{\textbf{R}}^\textbf{c}}_{\bar{Q}\bar{q}}$. The Eq.~\eqref{eq:Vspintesto} is obtained if we replace $\kappa_{ij}^{\textbf{R}^\textbf{c}}\mapsto\frac{1}{2}\mathcal{K}_{ij}(\textbf{R}^\textbf{c})$.

We show the detailed calculation for $S=1$, which is the most complete, and subsequently the cases $S=0$ and $S=2$.

\subsection{\texorpdfstring{$S=1$}{S=1}}
\label{sec:S=1}
The spin corrections are obtained by the diagonalization of the potential \eqref{eq:Vspin}, so we choose a starting basis and express the potential in matrix form.

In the color space, the tetraquark is represented by the ket $|(c\bar{c})^\textbf{8}(q\bar{q})^\textbf{8}\rangle$, which can be rearranged using the Fierz relations
\begin{equation}
    |(c\bar{c})^{\mathbf{8}}(q\bar{q})^{\mathbf{8}}\rangle=\sqrt{\frac{2}{3}}|(cq)^{\boldsymbol{\bar{3}}}(\bar{c}\bar{q})^{\boldsymbol{3}}\rangle-\sqrt{\frac{1}{3}}|(cq)^{\boldsymbol{6}}(\bar{c}\bar{q})^{\boldsymbol{\bar{6}}}\rangle
    \label{eq:fierzAapp}
\end{equation}
\begin{equation}
    |(c\bar{c})^{\mathbf{8}}(q\bar{q})^{\mathbf{8}}\rangle=\sqrt{\frac{8}{9}}|(c\bar{q})^{\boldsymbol{1}}(\bar{c}q)^{\boldsymbol{1}}\rangle-\sqrt{\frac{1}{9}}|(c\bar{q})^{\boldsymbol{8}}(\bar{c}q)^{\boldsymbol{8}}\rangle
    \label{eq:fierzBapp}
\end{equation}
In the spin space, we have three possible combinations yielding a spin 1 state
\begin{equation}
    |(c\bar{c})_0(q\bar{q})_1\rangle,\,\,|(c\bar{c})_1(q\bar{q})_0\rangle,\,\,|(c\bar{c})_1(q\bar{q})_1\rangle^{\phantom{\dagger}}_1
    \label{eq:spin_states}
\end{equation}
Regarding charge conjugation
\begin{equation}
    C(q\bar{q})_0=+(q\bar{q})_0,\quad\quad C(q\bar{q})_1=-(q\bar{q})_1
\end{equation}
The states $|(c\bar{c})_0(q\bar{q})_1\rangle$ and $|(c\bar{c})_1(q\bar{q})_0\rangle$ have $C=-1$, whereas the state $|(c\bar{c})_1(q\bar{q})_1\rangle^{\phantom{\dagger}}_1$ has $C=+1$. For completeness, we will present calculations for all states.

The basis \eqref{eq:spin_states} as it stands is not convenient for calculations because the spin of the $c(\bar{c})-q$ pairs is not explicit. We need to recouple the four spins to rewrite it, for example, as:
\begin{equation}
    |(c\bar{c})_0(q\bar{q})_1\rangle=A|(cq)_1(\bar{c}\bar{q})_0\rangle+B|(cq)_0(\bar{c}\bar{q})_1\rangle+C|(cq)_1(\bar{c}\bar{q})_1\rangle^{\phantom{\dagger}}_1
\end{equation}
We use the 9j-Wigner symbols to perform this change of basis. We refer to \cite{9jwigner} for a complete mathematical treatment of the argument.

Let $|S_1S_2(S_{12})S_3S_4(S_{34})S\rangle$ be the initial state, where $S_i$ is the spin of the $i$-th particle and $S$ is the total spin of the system, then the following relations hold
\begin{multline}
    \langle S_1S_2(S_{12})S_3S_4(S_{34})S|S_1S_3(S_{13})S_2S_4(S_{24})S\rangle=\\
    \sqrt{(2S_{12}+1)(2S_{13}+1)(2S_{24}+1)(2S_{34}+1)}
\left\{
    \begin{matrix}
        S_1 &S_2 & S_{12} \\
        S_3 & S_4 & S_{34} \\
        S_{13} & S_{24} & S
    \end{matrix}
\right\}
\label{eq:9j1}
\end{multline}
\begin{multline}
    \langle S_1S_2(S_{12})S_3S_4(S_{34})S|S_1S_4(S_{14})S_2S_3(S_{23})S\rangle=\\
    (-1)^{S_3+S_4-S_{34}}\sqrt{(2S_{12}+1)(2S_{13}+1)}\sqrt{(2S_{14}+1)(2S_{23}+1)}
\left\{
    \begin{matrix}
        S_1 &S_2 & S_{12} \\
        S_3 & S_4 & S_{34} \\
        S_{14} & S_{23} & S
    \end{matrix}
\right\}
\end{multline}
The symbols enclosed in curly brackets are called 9j-Wigner symbols and can be expressed as linear combinations of Clebsch-Gordan coefficients \cite{9jwigner}. For our purposes, the explicit expression is not necessary because they can be computed using \texttt{Wolfram Mathematica} \cite{Mathematica9j}. For example, if we want to determine the spin of the $c(\bar{c})-q(\bar{q})$ pairs from the state $|(c\bar{c})_0(q\bar{q})_1\rangle$, the following symbols are needed (to match the previous notation $S_1=S_c,\,S_2=S_{\bar{c}},\,S_3=S_q,\,S_4=S_{\bar{q}}$)
\begin{equation}
    \left\{
    \begin{matrix}
        \frac{1}{2} &\frac{1}{2} & 0 \\
        \frac{1}{2} & \frac{1}{2} & 1 \\
        1 & 0 & 1
    \end{matrix}
\right\}=-\frac{1}{6}\quad\quad
    \left\{
    \begin{matrix}
        \frac{1}{2} &\frac{1}{2} & 0 \\
        \frac{1}{2} & \frac{1}{2} & 1 \\
        0 & 1 & 1
    \end{matrix}
\right\}=\frac{1}{6}\quad\quad
    \left\{
    \begin{matrix}
        \frac{1}{2} &\frac{1}{2} & 0 \\
        \frac{1}{2} & \frac{1}{2} & 1 \\
        1 & 1 & 1
    \end{matrix}
\right\}=\frac{1}{3\sqrt{6}}
\end{equation}
from which we obtain
\begin{equation}
    |(c\bar{c})_0(q\bar{q})_1\rangle=-\frac{1}{2}|(cq)_1(\bar{c}\bar{q})_0\rangle+\frac{1}{2}|(cq)_0(\bar{c}\bar{q})_1\rangle+\frac{1}{\sqrt{2}}|(cq)_1(\bar{c}\bar{q})_1\rangle^{\phantom{\dagger}}_1
\end{equation}
Repeating the same process for the other two kets in \eqref{eq:spin_states}
\begin{equation}
    |(c\bar{c})_1(q\bar{q})_0\rangle=\frac{1}{2}|(cq)_1(\bar{c}\bar{q})_0\rangle-\frac{1}{2}|(cq)_0(\bar{c}\bar{q})_1\rangle+\frac{1}{\sqrt{2}}|(cq)_1(\bar{c}\bar{q})_1\rangle^{\phantom{\dagger}}_1
\end{equation}
\begin{equation}
    |(c\bar{c})_1(q\bar{q})_1\rangle^{\phantom{\dagger}}_1=\frac{1}{\sqrt{2}}|(cq)_0(\bar{c}\bar{q})_1\rangle+\frac{1}{\sqrt{2}}|(cq)_1(\bar{c}\bar{q})_0\rangle
    \label{eq:esempioccqq11}
\end{equation}

At this point, we can explicitly write the basis formed by spin 1 states, including color
\begin{equation}
 B_S=\left\{|(c\bar{c})^{\boldsymbol{8}}_0(q\bar{q})^{\boldsymbol{8}}_1\rangle^{\phantom{\dagger}}_1,\,\,|(c\bar{c})^{\boldsymbol{8}}_1(q\bar{q})^{\boldsymbol{8}}_0\rangle^{\phantom{\dagger}}_1,\,\,|(c\bar{c})^{\boldsymbol{8}}_1(q\bar{q})^{\boldsymbol{8}}_1\rangle^{\phantom{\dagger}}_1\right\}
\end{equation}
and depending on the interaction term to be studied, one can use one of the previously written relations (Fierz or 9-j symbols) to rewrite the basis in such a way that the total spin of the pair, which also appears in the spin operator, is explicit. For example, to determine the matrix element of the state $|(c\bar{c})^{\boldsymbol{8}}_1(q\bar{q})^{\boldsymbol{8}}_1\rangle^{\phantom{\dagger}}_1$ with the spin operators $\boldsymbol{S}_{c}\cdot\boldsymbol{S}_{q}$ (and charge conjugate), we use Eq.~\eqref{eq:fierzAapp} and \eqref{eq:esempioccqq11} to rewrite the state as
\begin{equation}
     |(c\bar{c})^{\boldsymbol{8}}_1(q\bar{q})^{\boldsymbol{8}}_1\rangle^{\phantom{\dagger}}_1=\frac{1}{\sqrt{3}}|(cq)^{\boldsymbol{\bar{3}}}_0(\bar{c}\bar{q})^{\boldsymbol{3}}_1\rangle-\frac{1}{\sqrt{6}}|(cq)^{\boldsymbol{6}}_0(\bar{c}\bar{q})^{\boldsymbol{\bar{6}}}_1\rangle+\frac{1}{\sqrt{3}}|(cq)^{\boldsymbol{\bar{3}}}_1(\bar{c}\bar{q})^{\boldsymbol{3}}_0\rangle-\frac{1}{\sqrt{6}}|(cq)^{\boldsymbol{6}}_1(\bar{c}\bar{q})^{\boldsymbol{\bar{6}}}_0\rangle
\end{equation}
and similarly for the other two.

To carry out all the computations, it is convenient (but not necessary) to switch to the basis
\begin{equation}
    B_S^\prime=\left\{\frac{1}{\sqrt{2}}\left(|(c\bar{c})^{\boldsymbol{8}}_0(q\bar{q})^{\boldsymbol{8}}_1\rangle-|(c\bar{c})^{\boldsymbol{8}}_1(q\bar{q})^{\boldsymbol{8}}_0\rangle\right),\frac{1}{\sqrt{2}}\left(|(c\bar{c})^{\boldsymbol{8}}_0(q\bar{q})^{\boldsymbol{8}}_1\rangle+|(c\bar{c})^{\boldsymbol{8}}_1(q\bar{q})^{\boldsymbol{8}}_0\rangle\right),|(c\bar{c})_1^{\boldsymbol{8}}(q\bar{q})_1^{\boldsymbol{8}}\rangle^{\phantom{\dagger}}_1\right\}
\end{equation}
and remember that since \eqref{eq:Vspin} is invariant under charge conjugation, the matrix elements between the first two states and the last one are always zero. The matrix representing $V_{q\bar{q}}^{\text{spin}}$ is a block matrix of dimensions $2\times 2$ ($C=-1$) and $1\times 1$ ($C=+1$). In particular, this last element is what we defined as $V_{q\bar{q}}^{\text{spin}}(1^{++})$ in Eq.~\eqref{eq:Vspinridotta}, which after tedious (but not complicated) steps can be shown to be
\begin{equation}
    V_{q\bar{q}}^{\text{spin}}(1^{++})=\;_1\langle(c\bar{c})_1^{\boldsymbol{8}}(q\bar{q})_1^{\boldsymbol{8}}| V_{q\bar{q}}^{\text{spin}}|(c\bar{c})_1^{\boldsymbol{8}}(q\bar{q})_1^{\boldsymbol{8}}\rangle^{\phantom{\dagger}}_1=-\frac{2}{3}\kappa_{cq}^{\boldsymbol{\bar{3}}}-\frac{1}{3}\kappa_{cq}^{\boldsymbol{6}}+\frac{1}{2}\kappa_{q\bar{q}}^{\boldsymbol{8}}-\frac{8}{9}\kappa_{\bar{c}q}^{\boldsymbol{1}}-\frac{1}{9}\kappa_{\bar{c}q}^{\boldsymbol{8}}
    \label{eq:Vspinp}
\end{equation}
The Eq.~\eqref{eq:Vspinridotta} is matched if we replace $(\kappa_{ij})^{\textbf{R}^\textbf{c}}\mapsto \mathcal{K}_{ij}(\textbf{R}^\textbf{c})/2$ and we use the relation between the quadratic Casimirs of SU(3).

For the subspace with negative charge conjugation, the potentials are the eigenvalues of the matrix
\begin{equation}
    \begin{pmatrix}
        \langle -|V_{q\bar{q}}^{\text{spin}}|-\rangle & \langle -|V_{q\bar{q}}^{\text{spin}}|+\rangle\\
        \langle -|V_{q\bar{q}}^{\text{spin}}|+\rangle & \langle +|V_{q\bar{q}}^{\text{spin}}|+\rangle
    \end{pmatrix}
\end{equation}
where 
\begin{eqnarray}
    |-\rangle&=&\frac{1}{\sqrt{2}}\left(|(c\bar{c})^{\boldsymbol{8}}_0(q\bar{q})^{\boldsymbol{8}}_1\rangle-|(c\bar{c})^{\boldsymbol{8}}_1(q\bar{q})^{\boldsymbol{8}}_0\rangle\right)\\
    |+\rangle&=&\frac{1}{\sqrt{2}}\left(|(c\bar{c})^{\boldsymbol{8}}_0(q\bar{q})^{\boldsymbol{8}}_1\rangle+|(c\bar{c})^{\boldsymbol{8}}_1(q\bar{q})^{\boldsymbol{8}}_0\rangle\right)
\end{eqnarray}
The explicit forms of the eigenvalues are
\begin{equation}
    V_{q\bar{q}}^{\text{spin},\pm}(1^{+-})=\pm\frac{1}{9}\sqrt{\left[6\kappa_{cq}^{\bar{\bm 3}}+3\kappa_{cq}^{\bm 6}-8\kappa_{\bar{c}q}^{\bm 1}-\kappa_{\bar{c}q}^{\bm 8}\right]^2+\left[9\kappa_{q\bar{q}}^{\bm 8}\right]^2}-\frac{1}{2}\kappa_{q\bar{q}}^{\bm 8}
    \label{eq:V_-spin}
\end{equation}
The Eq.~\eqref{eq:V_qq_neg} is given by the same substitution as before.

\subsection{\texorpdfstring{$S=0$}{S=0}}
The color quantum numbers are the same as in Eqs.~\eqref{eq:fierzAapp} and \eqref{eq:fierzBapp}, while the spin configurations that give $S=0$ are
\begin{equation}
    |(c\bar{c})_0(q\bar{q})_0\rangle^{\phantom{\dagger}}_0
\end{equation}
and
\begin{equation}
    |(c\bar{c})_1(q\bar{q})_1\rangle^{\phantom{\dagger}}_0
\end{equation}
Both have $C=+1$, so the matrix is not diagonal in this basis.

Using the 9-j symbols, we can recast these two states to highlight the couplings between heavy and light quarks (see Sec.~\ref{sec:S=1})
\begin{equation}
    |(c\bar{c})_0(q\bar{q})_0\rangle^{\phantom{\dagger}}_0=\frac{1}{2}|(cq)_0(\bar{c}\bar{q})_0\rangle^{\phantom{\dagger}}_0+\frac{\sqrt{3}}{2}|(cq)_1(\bar{c}\bar{q})_1\rangle^{\phantom{\dagger}}_0=-\frac{1}{2}|(\bar{c}q)_0(c\bar{q})_0\rangle^{\phantom{\dagger}}_0-\frac{\sqrt{3}}{2}|(\bar{c}q)_1(c\bar{q})_1\rangle^{\phantom{\dagger}}_0
\end{equation}
\begin{equation}
    |(c\bar{c})_1(q\bar{q})_1\rangle^{\phantom{\dagger}}_0=\frac{\sqrt{3}}{2}|(cq)_0(\bar{c}\bar{q})_0\rangle^{\phantom{\dagger}}_0-\frac{1}{2}|(cq)_1(\bar{c}\bar{q})_1\rangle^{\phantom{\dagger}}_0=+\frac{\sqrt{3}}{2}|(\bar{c}q)_0(c\bar{q})_0\rangle^{\phantom{\dagger}}_0-\frac{1}{2}|(\bar{c}q)_1(c\bar{q})_1\rangle^{\phantom{\dagger}}_0   
\end{equation}
The orthogonality between the two initial states is clearly preserved in the various relations. 

For simplicity, we define
\begin{eqnarray}
    |0\rangle&=&|(c\bar{c})_0^{\bm{8}}(q\bar{q})_0^{\bm{8}}\rangle\\ 
    |1\rangle&=&|(c\bar{c})_1^{\bm{8}}(q\bar{q})_1^{\bm{8}}\rangle^{\phantom{\dagger}}_0
\end{eqnarray}
Hence, the matrix that describes the spin interaction is
\begin{equation}
    \begin{pmatrix}
        \langle 0|V_{q\bar{q}}^{\text{spin}}|0\rangle & \langle 0|V_{q\bar{q}}^{\text{spin}}|1\rangle\\
        \langle 1|V_{q\bar{q}}^{\text{spin}}|0\rangle & \langle 1|V_{q\bar{q}}^{\text{spin}}|1\rangle
    \end{pmatrix}
\end{equation}
and it can be shown that
\begin{eqnarray}
    \langle0|V_{q\bar{q}}^{\text{spin}}|0\rangle&=&-\frac{3}{2}\kappa_{q\bar{q}}^{\bm{8}}\\
    \langle1|V_{q\bar{q}}^{\text{spin}}|1\rangle&=&-\frac{4}{3}\kappa_{cq}^{\bm{\bar{3}}}-\frac{2}{3}\kappa_{cq}^{\bm{6}}+\frac{1}{2}\kappa_{q\bar{q}}^{\bm{8}}-\frac{2}{9}\kappa_{\bar{c}q}^{\bm{8}}-\frac{16}{9}\kappa_{\bar{c}q}^{\bm{1}}\\
    \langle0|V_{q\bar{q}}^{\text{spin}}|1\rangle&=&-\frac{2}{\sqrt{3}}\kappa_{cq}^{\bm{\bar{3}}}-\frac{1}{\sqrt{3}}\kappa_{cq}^{\bm{6}}+\frac{\sqrt{3}}{9}\kappa_{\bar{c}q}^{\bm{ 8}}+\frac{8\sqrt{3}}{9}\kappa_{\bar{c}q}^{\bm{1}}
\end{eqnarray}
After calculating the eigenvalues of the matrix, replacing $\kappa_{ij}$ with $\frac{1}{2}\mathcal{K}_{ij}$, and using the quadratic Casimir relations of SU(3), the two potentials $V_{q\bar{q}}^{\text{spin},\pm}(0^{++})$ in \eqref{eq:Spin0} are obtained.

\subsection{\texorpdfstring{$S=2$}{S=2}}
This case is the most straightforward because the only color-spin combination is
\begin{equation}
    |(c\bar{c})^{\bm{8}}_1(q\bar{q})^{\bm{8}}_1\rangle^{\phantom{\dagger}}_2
\end{equation}
which has $C=+1$ and moreover its decomposition in the spin space is trivial since each possible pair must have spin 1. Therefore,
\begin{equation}
    2\kappa_{ij}^{\textbf{R}^\textbf{c}}\,\bm{S}_i\cdot\bm{S}_j=\frac{1}{2}\kappa_{ij}^{\textbf{R}^\textbf{c}}\quad \forall i,j
\end{equation}
from which we get (using Eqs.~\eqref{eq:fierzAapp} and \eqref{eq:fierzBapp})
\begin{equation}
    V_{q\bar{q}}^{\text{spin}}(2^{++})=\,_2\langle(c\bar{c})^{\bm{8}}_1(q\bar{q})^{\bm{8}}_1|V_{q\bar{q}}^{\text{spin}}|(c\bar{c})^{\bm{8}}_1(q\bar{q})^{\bm{8}}_1\rangle^{\phantom{\dagger}}_2=\frac{2}{3}\kappa_{cq}^{\bm{\bar{3}}}+\frac{1}{3}\kappa_{cq}^{\bm{6}}+\frac{1}{9}\kappa_{\bar{c}q}^{\bm{8}}+\frac{8}{9}\kappa_{\bar{c}q}^{\bm{1}}+\frac{1}{2}\kappa_{q\bar{q}}^{\bm{8}}
\end{equation}
which coincides with Eq.~\eqref{eq:Spin2} after the usual substitution.

\newpage

\section{\texorpdfstring{$\delta$}{delta} and \texorpdfstring{$\mathcal{K}(\textbf{R}^\textbf{c})$}{K} matrix elements}
\label{app:K}
\noindent We provide explicit expressions for the functions $K_{ij}(A;\textbf{R}^\textbf{c}),\,K_{ij}(B;\textbf{R}^\textbf{c}),\,K_{ij}(A,B;\textbf{R}^\textbf{c})$.

We will use the function $S_{AB}(R)$
\begin{equation}
\begin{split}
    S_{AB}(R)&=\int_{\boldsymbol{\xi}}\psi_A(\boldsymbol{\xi})\psi_B(\boldsymbol{\xi}-\boldsymbol{R})=\int_{\boldsymbol{\eta}}\psi_A(\boldsymbol{\eta})\psi_B(\boldsymbol{\eta}+\boldsymbol{R})=\\&=\frac{8\sqrt{A^3B^3}}{R(A^2-B^2)^2}\left(R\left(B e^{-A R}+A e^{-B R}\right)+\frac{4 A B}{A^2-B^2}\left(e^{-AR}-e^{-BR}\right)\right)
\end{split}
\end{equation}
and we also introduce 
\begin{equation}
\begin{split}
    L^A(R)&=\int_{\boldsymbol{\xi}}\psi_A(\boldsymbol{\xi})\psi_A(\boldsymbol{\xi}-\boldsymbol{R})=\int_{\boldsymbol{\eta}}\psi_A(\boldsymbol{\eta})\psi_A(\boldsymbol{\eta}+\boldsymbol{R})=\lim_{B\to A}S_{AB}(R)=\\&=e^{-AR}\left(1+AR+\frac{1}{3}A^2R^2\right)
\end{split}
\end{equation}
The matrix elements with Dirac deltas are ($\boldsymbol{\xi}^\prime=\boldsymbol{\xi}-\boldsymbol{R}$ and $\boldsymbol{\eta}^\prime=\boldsymbol{\eta}+\boldsymbol{R}$)
\begin{equation}
    \langle\psi_A(\boldsymbol{\xi})\psi_A(\boldsymbol{\eta})|\delta(\boldsymbol{\xi}-\boldsymbol{R})|\psi_A(\boldsymbol{\xi})\psi_A(\boldsymbol{\eta})\rangle=\langle\psi_A(\boldsymbol{\xi})\psi_A(\boldsymbol{\eta})|\delta(\boldsymbol{\eta}+\boldsymbol{R})|\psi_A(\boldsymbol{\xi})\psi_A(\boldsymbol{\eta})\rangle=\psi^2_A(\boldsymbol{R}) 
\end{equation}
\begin{align}
&  \langle\psi_A(\boldsymbol{\xi})\psi_A(\boldsymbol{\eta})|\delta(\boldsymbol{\xi})|\psi_A(\boldsymbol{\xi})\psi_A(\boldsymbol{\eta})\rangle=\langle\psi_A(\boldsymbol{\xi})\psi_A(\boldsymbol{\eta})|\delta(\boldsymbol{\eta})|\psi_A(\boldsymbol{\xi})\psi_A(\boldsymbol{\eta})\rangle=\psi_A^2(0)\\
    &\langle\psi_A(\boldsymbol{\xi})\psi_A(\boldsymbol{\eta})|\delta(\boldsymbol{\xi}-\boldsymbol{\eta}-\boldsymbol{R})|\psi_A(\boldsymbol{\xi})\psi_A(\boldsymbol{\eta})\rangle=\int_{\boldsymbol{\xi}}\psi_A^2(\boldsymbol{\xi})\,\psi^2_A(\boldsymbol{\xi}-\boldsymbol{R})=\frac{A^3}{8\pi}L^{2A}(R)\\
&    \langle\psi_B(\boldsymbol{\xi}^\prime)\psi_B(\boldsymbol{\eta}^\prime)|\delta(\boldsymbol{\xi}-\boldsymbol{R})|\psi_B(\boldsymbol{\xi}^\prime)\psi_B(\boldsymbol{\eta}^\prime)\rangle=\langle\psi_B(\boldsymbol{\xi}^\prime)\psi_B(\boldsymbol{\eta}^\prime)|\delta(\boldsymbol{\eta}+\boldsymbol{R})|\psi_B(\boldsymbol{\xi}^\prime)\psi_B(\boldsymbol{\eta}^\prime)\rangle=\psi_B^2(0)\\
&    \langle\psi_B(\boldsymbol{\xi}^\prime)\psi_B(\boldsymbol{\eta}^\prime)|\delta(\boldsymbol{\xi})|\psi_B(\boldsymbol{\xi}^\prime)\psi_B(\boldsymbol{\eta}^\prime)\rangle=\langle\psi_B(\boldsymbol{\xi}^\prime)\psi_B(\boldsymbol{\eta}^\prime)|\delta(\boldsymbol{\eta})|\psi_B(\boldsymbol{\xi}^\prime)\psi_B(\boldsymbol{\eta}^\prime)\rangle=\psi_B^2(\boldsymbol{R})\\
&    \langle\psi_B(\boldsymbol{\xi}^\prime)\psi_B(\boldsymbol{\eta}^\prime)|\delta(\boldsymbol{\xi}-\boldsymbol{\eta}-\boldsymbol{R})|\psi_B(\boldsymbol{\xi}^\prime)\psi_B(\boldsymbol{\eta}^\prime)\rangle=\int_{\boldsymbol{\xi}}\psi_B^2(\boldsymbol{\xi})\psi_B^2(\boldsymbol{\xi}-\boldsymbol{R})=\frac{B^3}{8\pi}L^{2B}(R)
\end{align}
and
\begin{align}
    \langle\psi_B(\boldsymbol{\xi}^\prime)\psi_B(\boldsymbol{\eta}^\prime)|\delta(\boldsymbol{\xi}-\boldsymbol{R})|\psi_A(\boldsymbol{\xi})\psi_A(\boldsymbol{\eta})\rangle&=\langle\psi_B(\boldsymbol{\xi}^\prime)\psi_B(\boldsymbol{\eta}^\prime)|\delta(\boldsymbol{\eta}+\boldsymbol{R})|\psi_A(\boldsymbol{\xi})\psi_A(\boldsymbol{\eta})\rangle=\notag\\&=S_{AB}(R)\,\psi_B(0)\psi_A(\boldsymbol{R})\\
    \langle\psi_B(\boldsymbol{\xi}^\prime)\psi_B(\boldsymbol{\eta}^\prime)|\delta(\boldsymbol{\xi})|\psi_A(\boldsymbol{\xi})\psi_A(\boldsymbol{\eta})\rangle&=\langle\psi_B(\boldsymbol{\xi}^\prime)\psi_B(\boldsymbol{\eta}^\prime)|\delta(\boldsymbol{\eta})|\psi_A(\boldsymbol{\xi})\psi_A(\boldsymbol{\eta})\rangle=\notag\\&=S_{AB}(R)\,\psi_B(\boldsymbol{R})\psi_A(0)\\
    \langle\psi_B(\boldsymbol{\xi}^\prime)\psi_B(\boldsymbol{\eta}^\prime)|\delta(\boldsymbol{\xi}-\boldsymbol{\eta}-\boldsymbol{R})|\psi_A(\boldsymbol{\xi})\psi_A(\boldsymbol{\eta})\rangle&=\int_{\boldsymbol{\xi}}\psi_B(\boldsymbol{\xi}-\boldsymbol{R})\psi_B(\boldsymbol{\xi})\psi_A(\boldsymbol{\xi})\psi_A(\boldsymbol{\xi}-\boldsymbol{R})=\notag\\
&=\frac{A^3B^3}{\pi(A+B)^3}L^{A+B}(R)
\end{align}
From these elements, we obtain the explicit expressions for $K_{ij}$ ($\boldsymbol{\xi}^\prime=\boldsymbol{\xi}-\bm{R}$ and $\boldsymbol{\eta}^\prime=\boldsymbol{\eta}+\bm{R}$)
\begin{equation}
    \mathcal{K}_{ij}(\textbf{R}_{ij}^c)=-\frac{\lambda_{ij}}{m_im_j}\frac{8\pi}{3}\alpha_s\,\delta^{(3)}(\boldsymbol{r}_i-\boldsymbol{r}_j)\quad \lambda_{q\bar{q}}(\mathbf{8})=+\frac{1}{6}\quad \lambda_{cq}(\bar{\boldsymbol{3}})=-\frac{2}{3}\quad \lambda_{c\bar{q}}(\boldsymbol{1})=-\frac{4}{3}
\end{equation}
\begin{align}
    &K_{c\bar{q}}(A;\textbf{1})=\langle\psi_A(\boldsymbol{\xi})\psi_A(\boldsymbol{\eta})| \mathcal{K}_{c\bar{q}}(\textbf{1})|\psi_A(\boldsymbol{\xi})\psi_A(\boldsymbol{\eta})\rangle=\frac{4}{3M_c\,m_q}\alpha_s\frac{8\pi}{3}\psi_A^2(\boldsymbol{R})\\
    &K_{cq}(A;\mathbf{\bar{3}})=\langle\psi_A(\boldsymbol{\xi})\psi_A(\boldsymbol{\eta})| \mathcal{K}_{cq}(\mathbf{\bar{3}})|\psi_A(\boldsymbol{\xi})\psi_A(\boldsymbol{\eta})\rangle=\frac{2}{3M_c\,m_q}\alpha_s\frac{8\pi}{3}\psi^2_A(0)\\
    &K_{q\bar{q}}(A;\textbf{8})=\langle\psi_A(\boldsymbol{\xi})\psi_A(\boldsymbol{\eta})| \mathcal{K}_{q\bar{q}}(\textbf{8})|\psi_A(\boldsymbol{\xi})\psi_A(\boldsymbol{\eta})\rangle=-\frac{1}{6\,m_q^2}\alpha_s\frac{8\pi}{3}\frac{A^3}{8\pi}L^{2A}(R)\\
    &K_{c\bar{q}}(B;\textbf{1})=\langle\psi_B(\boldsymbol{\xi}^\prime)\psi_B(\boldsymbol{\eta}^\prime)| \mathcal{K}_{c\bar{q}}(\textbf{1})|\psi_B(\boldsymbol{\xi}^\prime)\psi_B(\boldsymbol{\eta}^\prime)\rangle=\frac{4}{3M_c\,m_q}\alpha_s\frac{8\pi}{3}\psi_B^2(0)\\
    &K_{cq}(B;\mathbf{\bar{3}})=\langle\psi_B(\boldsymbol{\xi}^\prime)\psi_B(\boldsymbol{\eta}^\prime)|\mathcal{K}_{cq}(\mathbf{\bar{3}})|\psi_B(\boldsymbol{\xi}^\prime)\psi_B(\boldsymbol{\eta}^\prime)\rangle=\frac{2}{3M_c\,m_q}\alpha_s\frac{8\pi}{3}\psi_B^2(\boldsymbol{R})\\
    &K_{q\bar{q}}(B;\textbf{8})=\langle\psi_B(\boldsymbol{\xi}^\prime)\psi_B(\boldsymbol{\eta}^\prime)| \mathcal{K}_{q\bar{q}}(\textbf{8})|\psi_B(\boldsymbol{\xi}^\prime)\psi_B(\boldsymbol{\eta}^\prime)\rangle=-\frac{1}{6\,m_q^2}\alpha_s\frac{8\pi}{3}\frac{B^3}{8\pi}L^{2B}(R)\\
    &K_{c\bar{q}}(A,B;\textbf{1})=\langle\psi_B(\boldsymbol{\xi}^\prime)\psi_B(\boldsymbol{\eta}^\prime)| \mathcal{K}_{c\bar{q}}(\textbf{1})|\psi_A(\boldsymbol{\xi})\psi_A(\boldsymbol{\eta})\rangle=\frac{4}{3M_c\,m_q}\alpha_s\frac{8\pi}{3}S_{AB}(R)\,\psi_B(0)\psi_A(\boldsymbol{R})\\
    &K_{cq}(A,B;\mathbf{\bar{3}})=\langle\psi_B(\boldsymbol{\xi}^\prime)\psi_B(\boldsymbol{\eta}^\prime)|\mathcal{K}_{cq}(\mathbf{\bar{3}})|\psi_A(\boldsymbol{\xi})\psi_A(\boldsymbol{\eta})\rangle=\frac{2}{3M_c\,m_q}\alpha_s\frac{8\pi}{3}S_{AB}(R)\,\psi_B(\boldsymbol{R})\psi_A(0)\\
    &K_{q\bar{q}}(A,B;\textbf{8})=\langle\psi_B(\boldsymbol{\xi}^\prime)\psi_B(\boldsymbol{\eta}^\prime)|\mathcal{K}_{q\bar{q}}(\textbf{8})|\psi_A(\boldsymbol{\xi})\psi_A(\boldsymbol{\eta})\rangle=-\frac{1}{6\,m_q^2}\alpha_s\frac{8\pi}{3}\frac{A^3B^3}{\pi(A+B)^3}L^{A+B}(R)
\end{align}

\newpage
\bibliographystyle{JHEP}
{\footnotesize
\bibliography{biblio}}

\end{document}